\documentclass[10pt]{IEEEtran}

\usepackage{amsmath,amssymb}

\usepackage{subfigure}
\usepackage{graphicx}
\usepackage{float}
\usepackage{caption}
\usepackage{slashbox}
\usepackage{longtable}
\usepackage{rotating}
\usepackage{multirow}
\usepackage{array}
\usepackage{bm}
\usepackage{algorithmic}

\usepackage[ruled,vlined,linesnumbered]{algorithm2e}
\usepackage{setspace}
\usepackage[usenames]{color}
\usepackage{cases}
\usepackage{CJK}
\usepackage{tabularx}
\usepackage{xcolor}
\usepackage{colortbl,booktabs}
\usepackage{amscd}
\usepackage{multirow}
\usepackage{bigdelim}
\usepackage{caption}
\usepackage{ragged2e}
\usepackage{indentfirst}
\usepackage{setspace}
\usepackage{epstopdf}
\usepackage{pifont}

\definecolor{mygray}{gray}{.9}
\definecolor{myblue}{RGB}{135,206,250}
\definecolor{mybluegray}{RGB}{119,136,153}

\begin{document}

\title{Blockchain for Data Sharing at the Network Edge: Trade-Off Between Capability and Security}

\author{Yixin~Li,
        Liang~Liang$^*$,
        Yunjian~Jia,~\IEEEmembership{Member,~IEEE,}
        Wanli~Wen,~\IEEEmembership{Member,~IEEE,}\\
        Chaowei~Tang,~\IEEEmembership{Member,~ACM,}
        and~Zhengchuan~Chen,~\IEEEmembership{Member,~IEEE}\vspace{-0.4cm}
\thanks{Y. Li, L. Liang, Y. Jia, W. Wen, C. Tang, and Z. Chen are with the School of Microelectronics and Communication
Engineering, Chongqing University, Chongqing 400044, China. E-mail: \{liyixin, liangliang, yunjian, wanli\_wen, cwtang, czc\}@cqu.edu.cn. (Corresponding author: Liang Liang.)}
\thanks{This work was supported by National Natural Science Foundation of China under Grant 62071075.}
}

\IEEEtitleabstractindextext{%
\begin{abstract}
\justifying
Blokchain is a promising technology to enable distributed and reliable data sharing at the network edge. The high security in blockchain is undoubtedly a critical factor for the network to handle important data item. On the other hand, according to the dilemma in blockchain, an overemphasis on distributed security will lead to poor transaction-processing capability, which limits the application of blockchain in data sharing scenarios with high-throughput and low-latency requirements. To enable demand-oriented distributed services, this paper investigates the relationship between capability and security in blockchain from the perspective of block propagation and forking problem. First, a Markov chain is introduced to analyze the gossiping-based block propagation among edge servers, which aims to derive block propagation delay and forking probability. Then, we study the impact of forking on blockchain capability and security metrics, in terms of transaction throughput, confirmation delay, fault tolerance, and the probability of malicious modification. The analytical results show that with the adjustment of block generation time or block size, transaction throughput improves at the sacrifice of fault tolerance, and vice versa. Meanwhile, the decline in security can be offset by adjusting confirmation threshold, at the cost of increasing confirmation delay. The analysis of capability-security trade-off can provide a theoretical guideline to manage blockchain performance based on the requirements of data sharing scenarios.

\end{abstract}

}

\maketitle

\IEEEdisplaynontitleabstractindextext

%
\IEEEpeerreviewmaketitle


%
%
%
%


\section{Introduction}

Edge computing extends cloud resources to the network edge by deploying geographically distributed edge servers, which provide efficient data processing services for edge devices. Nowadays, edge devices are equipped with advanced sensing technologies that have produced massive amounts of data belonging to different stakeholders. With the intermediary function of edge servers, the data in edge devices is expected to be processed and shared to enhance task collaborations, improve driving safety, and create new business models. Although the distributed services of edge computing can reduce the backbone network load and the single point of failure in cloud computing, the security and trust issues on edge servers are significant challenges, due to the lack of a technique for distributed coordination and transparent data processing \cite{5-RuizheYang}.

As a distributed ledger, blockchain comes with the characteristics of high security, trust-building, and traceability, which has the potential to establish a reliable and distributed data sharing platform. First, using a hash-based chain structure, the data stored in blockchain is immutable, unless the attacker obtains more than half of the consensus resource (e.g., hashrate or stake) to recast the chain. Second, blockchain can be an open system maintained by all stakeholders, and no one can secretly control the data without punishment. Furthermore, the data sharing history can be stored in a distributed manner for any user to trace and verify data source efficiently.


In recent years, the integration of blockchain and data sharing has attracted extensive attentions in the research community. The current researches mainly focus on architecture design \cite{6-MengShen}, consensus optimization \cite{6-YueqiangXu}, copyright protection \cite{4-YaodongHuang}, and reputation management \cite{6-Jiawen}, while the suitability between the basic performance of blockchain and the service requirements of data sharing has not been well investigated. According to the technical specification of 3GPP \cite{1-3GPP}, high throughput, low latency, and high security have been identified as the key performance requirements for future networks to handle different types of data. The high security of blockchain is undoubtedly a critical factor for the network to handle important data item, but an overemphasis on security will compromise decentralization or lead to poor transaction-processing capability (i.e., low transaction throughput and high confirmation delay), according to the trilemma that blockchain systems can only have two elements in decentralization, capability, and security. The poor capability limits the application of blockchain in high-traffic and real-time data sharing scenarios, such as smart cities and connected vehicles.

To enable demand-oriented distributed services, this paper investigates the relationship between capability and security in blockchain from the perspective of block propagation and forking problem. We start by introducing a Markov chain to study block propagation performance metrics, in terms of the increasing rate of informed servers, block propagation delay, and the probability of propagation failure. Based on block propagation performance, we determine forking probability and establish mathematical relationships between blockchain system parameters and capability-security metrics, in terms of transaction throughput, confirmation delay, fault tolerance, and the probability of malicious modification. The main contributions of this paper can be summarized as follows.

1) \begin{bfseries}Modeling:\end{bfseries} We model the block propagation process based on gossip protocol as a Markov chain, which captures the dynamic change of the number of informed servers (who have received the new block) over time and the impact of asynchronous block transmissions in the network.

2) \begin{bfseries}Metrics:\end{bfseries} We derive the closed-form expressions of block propagation performance metrics based on Markov chain, which act as a fundamental to determine forking probability. Then, we derive the expressions of capability and security metrics by analyzing the impact of forking on blockchain.

3) \begin{bfseries}Insights:\end{bfseries} We reveal the theoretical performance bounds and trade-offs of blockchain, which can serve as a guidance for data sharing applications.
\begin{itemize}
\item Bounds: With the exponential assumption of block transmission time, the lower bound of block propagation delay can be expressed as a logarithmic function with Euler's constant. Due to forking, transaction throughput and confirmation delay have upper and lower bounds respectively, which are determined by the network parameters of block propagation, i.e., the total number of servers, the number of selected servers, and network data rate.
\item Trade-offs: As block generation time decreases or block size increases, transaction throughput can approach its upper bound gradually, at the sacrifice of fault tolerance. Meanwhile, confirmation threshold can be increased to offset the decline in security, but it results in a higher confirmation delay. Based on the trade-offs, we can adjust system parameters to satisfy the required performance in data sharing, while capturing its adverse impact on the other performance in a quantitative manner.
\end{itemize}

The rest of the paper is organized as follows. Section II summarizes the related works on blockchain-based data sharing and theoretical modeling. Section III describes the block propagation process of gossip protocol and forking problem. Section IV introduces a Markov chain to model the block propagation process among edge servers. Based on the transition probabilities of Markov chain, the closed-form expressions of blockchain performance metrics are analyzed in Section V. Section VI conducts numerical experiments to evaluate the performance trade-offs and bounds of blockchain, and Section VII concludes the paper.

\section{Related Work}

\subsection{Blockchain Architecture for Data Sharing}

Since edge devices have limited computing and storage resources, early solutions suggested a blockchain architecture that considers edge servers as blockchain full nodes to maintain a full ledger and handle the data from edge devices \cite{7-JiawenKang}, \cite{7-LiangXiao}. However, this architecture requires all edge servers to store the same blockchain ledger, which has serious scalability issues on data volume and cannot provide demand-oriented data sharing services for different applications. For example, the road information in vehicle networks has location-related regional characteristics \cite{5-ZheYang}. To address this issue, the authors in \cite{4-HaoyeChai} proposed a hierarchical blockchain architecture, which divides the network into multiple consensus domains based on regional characteristics. The edge servers in a consensus domain maintain a sub-blockchain ledger for intra-domain data sharing, and region data centers maintain a global blockchain ledger for cross-domain data sharing. Based on the hierarchical architecture, the authors in \cite{6-DekeGuo} designed a data sharing protocol called Cuckoo Summary to achieve fast data localization. In \cite{5-JianChang}, the authors adopted a hierarchical architecture to enable cross-application data sharing.

The above works discussed the advantages of hierarchical blockchain architecture from the regional or cross-application characteristics for data sharing, while the impact of hierarchical architecture on blockchain performance has not been studied from a theoretical perspective. Compared with the flat architecture, hierarchical architecture gives an advantage to manage the capability-security trade-off of blockchain based on the performance requirement in each consensus domain. Meanwhile, hierarchical architecture has less edge servers and shorter block transmission times in a consensus domain, which accelerate block propagation. In view of this, we conduct a theoretical analysis to explain how a faster block propagation can reduce forking probability, and why a lower forking probability is beneficial to blockchain capability improvement with less compromise on security.

\subsection{Block Propagation Modeling and Analysis}

As a distributed ledger, blockchain relies on block propagation among full nodes to achieve information synchronization, which is a critical factor to determine overall system performance \cite{5-PinarOzisik}. Most of the prior works studied block propagation in Bitcoin network. Thousands of full nodes in Bitcoin lead to complex block propagation process, which can only be captured by network simulator or approximate model. For instance, the authors in \cite{2-ChristianDecker} adopted a network simulator to study the block propagation in Bitcoin, and shows that the number of informed nodes follows exponential growth and convergence behaviors over time. In \cite{5-Jelena}, the authors divided the whole block propagation process of Bitcoin into multiple generations. Each generation represented a single-hop block propagation from informed nodes to their neighbors. Then, the mean number of informed nodes in a given generation was derived based on the long-tail distribution of node connectivity. Similar to the idea in \cite{5-Jelena}, the authors in \cite{3-YahyaShahsavari} divided the block propagation process into multiple waves, and then calculated the number of informed nodes in any wave using a random graph model. The above models provided an approximate solution for block propagation process by assuming that all the nodes in each generation or wave can receive and forward blocks at the same time, namely that the block transmissions in the network are assumed to be synchronous.

In this work, we consider the block propagation in a consensus domain at the network edge, where edge servers work as blockchain full nodes. Since the number of edge servers in a consensus domain are much less than the full nodes in Bitcoin, it allows us to capture the impact of asynchronous block transmissions in the network using a Markov chain. The proposed model derives the mathematical relationship between block propagation delay and forking probability, which theoretically explains why blockchain capability-security trade-off results from forking.

\section{Preliminaries}


\subsection{Blockchain Architecture and Consensus Process}

As shown in Fig. 1(a), we consider a hierarchical blockchain architecture for data sharing, which consists of one global blockchain ledger and multiple sub-blockchain ledgers. Each sub-ledger is maintained by all the edge servers in a consensus domain, which is formed based on geographical locations or application scenarios. These sub-ledgers work as trusted platforms to handle and share the data among the edge devices within their own consensus domains. On the other hand, region data centers can request for the confirmed data in edge servers, and then generate a global ledger for cross-domain data sharing. In this architecture, region data centers do not involve the consensus process of sub-ledger. Each sub-ledger is functionally independent, which provides a distributed and robust data sharing service with resistance to the single point of failure.


Existing researches on blockchain-based data sharing mainly proposed two types of consensus mechanisms: proof-type and voting-type consensuses. Proof-type consensus is a variant of the original proof-of-work \cite{1-Nakamoto} in Bitcoin, which includes proof-of-collaboration \cite{7-ChenhanXu}, proof-of-knowledge \cite{6-JiananLi}, and proof-of-utility \cite{5-YueyueDai}. Voting-type consensus includes Byzantine fault tolerant \cite{4-JinHuaChen}, practical Byzantine fault tolerant \cite{6-JinmingShi}, and reputation-based Byzantine fault tolerance \cite{7-LiangYuan}. In this work, we focus on the performance trade-offs of proof-type consensus, which has better decentralization (no leader) and fault tolerance ($50\%$ consensus resources) than voting-type consensus \cite{4-YangXiao}. The main steps of the consensus process in a consensus domain can be summarized as follows:
1) Transaction collection: The edge servers collect the required data items (refer to transactions in blockchain) from edge devices or the cloud data center, and then include them into candidate blocks.
2) Block generation: All edge servers compete for generating a valid new block using a specified consensus resource, e.g., hashrate and stake.
3) Block propagation: The new block is propagated to the other servers using blockchain gossip protocol.
4) Block accumulation: The new block waits for the accumulation of subsequent blocks until reaching a confirmation threshold, which is considered a sufficient proof that the block and its transactions cannot be reversed.

\subsection{Block Propagation and Forking Problem}

\begin{figure}[t]
\setlength{\abovecaptionskip}{0.cm}
\setlength{\belowcaptionskip}{-0.cm}
\captionsetup{font={footnotesize}}
\centering
\subfigure[A hierarchical blockchain architecture for data sharing.]{
\includegraphics[width=8.5cm]{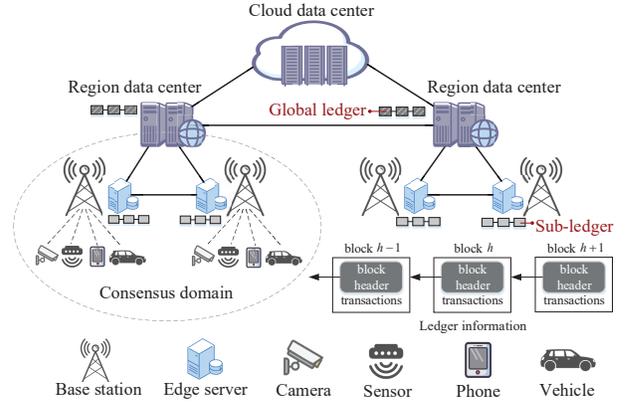}
}
\subfigure[Block propagation in a consensus domain and forking problem.]{
\includegraphics[width=7.5cm]{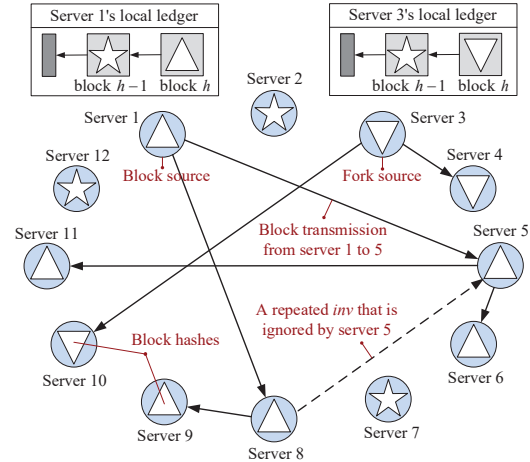}
}
\caption{{An illustration of blockchain architecture and block propagation.}}
\label{gossiping}
\end{figure}

In a consensus domain, all the edge servers work as blockchain full nodes to store a local copy of the ledger information. To full update a distributed ledger, the new block should be propagated to all full nodes. Since any node in a distributed environment might encounter Byzantine failure caused by the attacker, Bitcoin and most blockchain systems adopt a gossip protocol that randomly selects a given number of nodes to forward the new block \cite{2-ChristianDecker}, \cite{4-YangXiao}.

In this work, we study the gossiping-based block propagation in a consensus domain. Fig. \ref{gossiping}(b) visualizes a block propagation from edge server 1, where \ding{73}, $\vartriangle$, and $\triangledown$ represent block hashes. Note that the block hash can identify a block uniquely \cite{1-Antonopoulos}. Once a new block with hash $\vartriangle$ is generated, server 1 as a block source will randomly select a given number of servers in its consensus domain (two servers are selected as an example). To avoid redundant block transmission, server 1 performs a two-way handshake protocol: an \emph{inventory} (\emph{inv}) message is sent to all the selected servers, and then a \emph{getdata} message is sent back if the block is needed. After receiving a new block, the server will continue to forward it using gossiping with two-way handshake.

During block propagation, different servers may receive the new block with hash $\vartriangle$ at different times, due to multi-hop gossiping and the randomness of block transmission time in the network. Those servers, that have not received hash $\vartriangle$, keep on mining new blocks based on hash \ding{73}. Suppose server 3 generates another new block with hash $\triangledown$ before receiving $\vartriangle$. At this time, hashes $\vartriangle$ and $\triangledown$ are both calculated based on hash \ding{73}, and thus the two new blocks have the same height $h$ (the position in blockchain), causing a forking problem. When forking occurs, servers will mine the next block based on the hash that is received first, and the network cannot determine which block is valid due to the disagreement among servers.

Blockchain uses a longest-chain rule to address forking problem. To maximize its profit, a rational server should work on the longest chain when forking occurs, since the longest chain has the lowest probability to be orphaned. Based on the longest-chain rule, the servers that have a disagreement on blocks $h$ will wait for the propagation of block $h+1$. If the propagation of block $h+1$ does not incur new forks, all servers will follow an unique block $h+1$, and one of the blocks $h$ will be orphaned. The hash of an orphaned block will not be used to mine new blocks anymore, and thus the transactions included in it cannot reach confirmation threshold, which affects the overall transaction-processing capability of blockchain. Meanwhile, forking slows down the block accumulation on the longest chain by incurring parallel branches, which makes it easier for attacker to outpace the longest chain through deliberate forking and replace the confirmed transactions.

\begin{figure*}[t]
\setlength{\abovecaptionskip}{0.cm}
\setlength{\belowcaptionskip}{-0.2cm}
\captionsetup{font={footnotesize}}
\begin{center}
 \includegraphics[width=16cm]{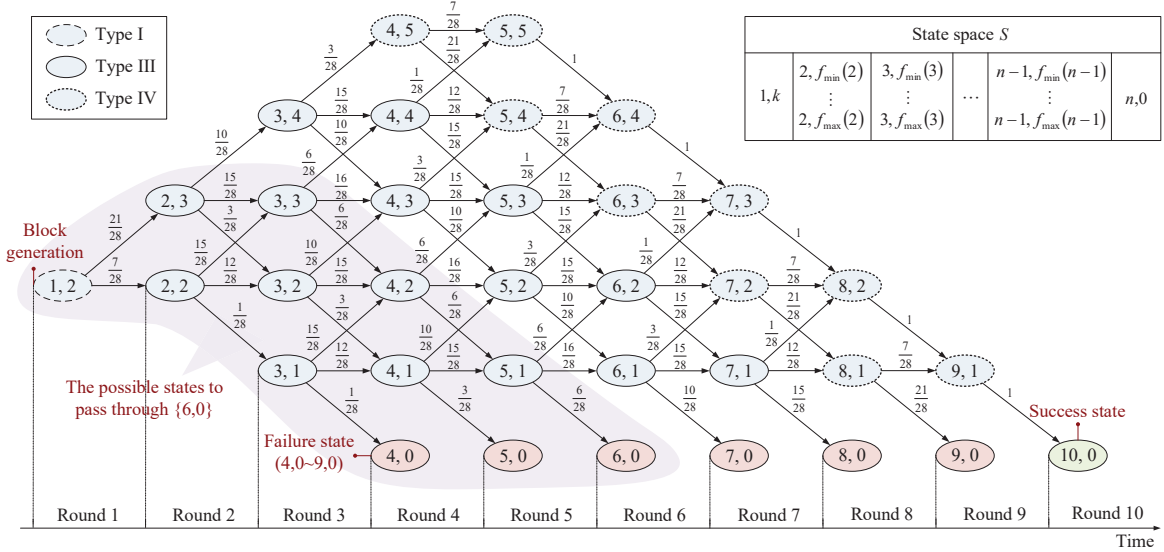}
 \end{center}
 \caption{An example of Markov chain for block propagation; when $n=10$ and $k=2$.}
\label{Markovfig}
\end{figure*}

\section{Block Propagation Modeling}

In this section, we model the block propagation process in a consensus domain as a Markov chain to provide a theoretical fundamental for the performance analysis of blockchain.

\subsection{Definitions and Assumptions}

We consider a consensus domain with $n$ edge servers that propagate a new block $h$ by randomly selecting $k$ servers. During block propagation, each server (except the block source) must pass through three stages: Initially, a server does not have any messages about block $h$, which is called an \emph{uninformed} server. Then, the server receives an \emph{inv} about block $h$ and accepts for a block transmission, which is called an \emph{engaged} server. After block transmission and verification, the server will agree with the validity of block $h$, which is called an \emph{informed} server. With the propagation of block $h$, the number of informed server will gradually increase from $1$ to $n$, and its increasing rate is affected by the number of engaged servers. To study the increasing rate of informed servers, we divide block propagation process into multiple rounds $r$, where $1\leq r \leq n$. Each round describes a competition among engaged servers for who can become an informed server first, namely that a round ends when the number of informed servers increased by $1$. Let $T_r$ denote the time interval from round $r$ to round $r+1$, and then the increasing rate of informed servers in round $r$ is $1/T_r$.

Based on the definition of round, $T_r$ can be interpreted as the time until one of the engaged servers wins the competition, which means the server should receive and verify block $h$ with the shortest amount of time. Since block verification time only takes few milliseconds \cite{5-Jelena}, we assume that it is negligible compared with the time to transmit a block ($1$ MB) over the network. Based on the assumption, $T_r$ is approximately equal to the shortest block transmission time in round $r$. We consider that edge servers are connected through wired backhaul, and the block transmission time $T_{m}$ in backhaul network can be modeled as an exponentially distributed random variable having expectation $E[T_{m}]$ proportional to block size \cite{3-Chen}. The expected block transmission time in backhaul network can be expressed as
\begin{equation}\label{Tb}
\begin{split}
E[T_{m}]=\frac{s_b}{\lambda_d},
\end{split}
\end{equation}
where $s_b$ is block size, and $\lambda_d$ is the data rate in backhaul links. Let $E_r$ denote the number of engaged servers at the beginning of round $r$, which indicates that the number of in-transit block $h$ is equal to $E_r$. Let $T_{m1}$, $T_{m2}$, ..., $T_{mE_r}$ be the block transmission time of engaged servers in round $r$, which are independent identically distributed exponential random variables having expectation ${s_b}/{\lambda_d}$. Recalling that $T_r$ is the shortest block transmission time in round $r$, so it is given by
\begin{equation}\label{Tr}
\begin{split}
T_{r}=\min\{T_{m1}, T_{m2}, ... , T_{mE_r}\}.
\end{split}
\end{equation}
Based on the property of exponential distribution \cite{1-Ross}, $T_{r}$ is an exponential random variable having expectation ${s_b}/({\lambda_d}E_r)$. To determine $T_{r}$, the key challenge is to analyze $E_r$, which is affected by the random selection in each round.


\subsection{Markov Chain for Block Propagation}

Let $U_r$, $E_r$, and $I_r$ denote the number of uninformed, engaged, and informed servers at the beginning of round $r$ respectively, which satisfy $U_r+E_r+I_r=n$. We model block propagation process as a two-dimensional Markov chain $\{I_r, E_r\}$ to capture the impact of random selection on the number of engaged servers in each round. The state space and the one-step transition probabilities of Markov chain will be analyzed for an arbitrary block propagation process with $n\geq2$ and $1\leq k \leq n-1$. An example for $n=10$ and $k=2$ is presented in Fig. \ref{Markovfig}. At the beginning of a round, a new informed server conducts a random selection of $k$ servers from a set of $n-2$ servers (with the exception of itself and the server that transmitted block to it). In the set of $n-2$ servers, there are $I_r+E_r-2$ servers that have received the \emph{inv} about block $h$, and $U_r$ servers without the \emph{inv}. The value of $E_{r+1}$ is determined by the outcome of random selection: if the new informed server selects $0$ element in $U_r$, $E_{r+1}=E_r-1$; if it selects $1$ element in $U_r$, $E_{r+1}=E_r$; it can select up to $k$ elements in $U_r$, and $E_{r+1}=E_r+k-1$. This means that $E_r$ can decrease by $1$ or increase by $k-1$ at most in each round, so it follows that $E_1=k$, $E_2\in[k, 2k-1]$, $E_3\in[k-1, 3k-2]$, $E_4\in[k-2, 4k-3]$, ..., $E_r\in[k-r+2, rk-r+1]$ ($2\leq r \leq n$). Note that round $1$ to round $2$ is a special case in which the informed server must select at least $1$ element in $U_r$ due to $I_r+E_r-2=k-1$. On the other hand, based on the definition of $E_r$, we have $E_r\geq0$ and $E_r+I_r\leq n$, and it follows that $E_r\in[0, n-r]$ (using $I_r=r$). So in summary, we can obtain
\begin{equation}\label{Er}
\begin{split}
E_r&\in[k-r+2, rk-r+1]\cap[0, n-r]\\
&\in[f_{\max}(r),f_{\min}(r)],
\end{split}
\end{equation}
where $f_{\max}(r)=\max\{k-r+2, 0\}$ and $f_{\min}(r)=\min\{rk-r+1, n-r\}$, $2\leq r \leq n$. Based on (\ref{Er}), the state space $S$ of Markov chain is shown in Fig. \ref{Markovfig}.

Since the outcome of random selection is affected by the value of $I_r+E_r-2$ and $U_r$, the Markov chain may contain four types of states that have different transition probabilities (except for success and failure states). The state $\{I_r, E_r\}$ in Markov chain is defined as
\begin{equation}\label{statetype}
\begin{split}
\begin{cases}
\text{Type I},~~~\text{if}~I_r+E_r-2<k\text{ and }U_r\geq k,\\
\text{Type II},~~\text{if}~I_r+E_r-2<k\text{ and }U_r< k,\\
\text{Type III},~\text{if}~I_r+E_r-2\geq k\text{ and }U_r\geq k,\\
\text{Type IV},~\text{if}~I_r+E_r-2\geq k\text{ and }U_r< k.
\end{cases}\\
\end{split}
\end{equation}
Type I state $\{I_r, E_r\}$ can transit to $k$ next states $\{I_{r+1}, E_{r+1}\}$ and satisfies $E_{r+1}\in[E_r, E_r+k-1]$. Type II state can transit to $U_r$ next states and satisfies $E_{r+1}\in[E_r, E_r+U_r-1]$. Type III state can transit to $k+1$ next states and satisfies $E_{r+1}\in[E_r-1, E_r+k-1]$. Type IV state can transit to $U_r+1$ next states and satisfies $E_{r+1}\in[E_r-1, E_r+U_r-1]$. Note that Type II state does not appear in Fig. \ref{Markovfig} since $I_r+E_r-2<k\text{ and }U_r< k$ are impossible when $n=10,~k=2$. Based on the types of state, the one-step transition probabilities can be expressed as
\begin{equation}\label{onestep}
\begin{split}
P\left\{ {I_r+1,E_r+e\mid I_r, E_r} \right\}=\frac{{C_{I_r+E_r-2}^{k-e-1} C_{U_r}^{e+1}}}{{C_{n-2}^{k}}},
\end{split}
\end{equation}
where $C_{n-2}^{k}$ denotes the number of combinations that selecting $k$ servers from $n-2$ servers at random. Meanwhile, type I state satisfies $e\in[0,k-1]$; type II state satisfies $e\in[0,U_r-1]$; type III state satisfies $e\in[-1,k-1]$; and type IV state satisfies $e\in[-1,U_r-1]$.


Now we should determine the value range of the four types of states, which are denoted by $S_{I}$, $S_{II}$, $S_{III}$, and $S_{IV}$.
According to $E_r\in[f_{\max}(r),f_{\min}(r)]$ in (\ref{Er}), we can know that $E_r$ has a lower bound $0$ and a upper bound $n-r$. Solving $k-r+2=0$ and $rk-r+1=n-r$ in (\ref{Er}), we obtain that $r=k+2$ and $r=\lceil\frac{n-1}{k}\rceil$ are respectively the first round that the Markov chain reaches the lower and upper bounds of $E_r$. Based on this, $S_{I}$ to $S_{IV}$ can be derived for an arbitrary block propagation process with $1\leq k \leq n-1$ as follows:

1) When $k\geq1$ and $k+2<\lceil\frac{n-1}{k}\rceil$, namely $1\leq k<\sqrt{n}-1$, the Markov chain reaches lower bound faster than upper bound. In this case, the state space can rewrite as
\begin{equation}\label{S1}
\begin{split}
S=
\begin{cases}
\{I_1=1,E_1=k\},\\
\{I_r\in[{2,k+1}],\underbrace{E_r\in[k-r+2,rk-r+1]}_{\text{has not reach bound yet}}\},\\
\{I_r\in[k+2,\lceil\frac{n-1}{k}\rceil-1],\underbrace{E_r\in[0,rk-r+1]}_{\text{reaches lower bound}}\},\\
\{I_r\in[\lceil\frac{n-1}{k}\rceil,n],\underbrace{E_r\in[0,n-r]}_{\text{reaches two bounds}}\},~~~~(I_r=r).
\end{cases}
\end{split}
\end{equation}
Accordingly, $S_{I}$ to $S_{IV}$ are given by
\begin{equation}\label{S1type}\small
\begin{split}
&S_{I}=\{I_1=1,E_1=k\},~S_{II}=\phi~(\text{empty set}),\\
&S_{III}=\\
&\begin{cases}
\{I_r\in[2,k],E_r\in[k-r+2,rk-r+1]\},\\
\{I_r\in[k+1,\lceil\frac{n-1}{k}\rceil-2],E_r\in[1,rk-r+1]\},\\
\{I_r\in[\lceil\frac{n-1}{k}\rceil-1,n-k-1],E_r\in[1,n-r-k]\},
\end{cases}\\
&S_{IV}=\\
&\begin{cases}
\{I_r=\lceil\frac{n-1}{k}\rceil-1,E_r\in[n-r-k+1,rk-r+1]\},\\
\{I_r\in[\lceil\frac{n-1}{k}\rceil,n-k-1],E_r\in[n-r-k+1,n-r]\},\\
\{I_r\in[n-k,n-1],E_r\in[1,n-r]\},
\end{cases}
\end{split}
\end{equation}


2) When $k+2\geq\lceil\frac{n-1}{k}\rceil$ and $k\leq \frac{n-2}{2}$, namely $\sqrt{n}-1\leq k\leq \frac{n-2}{2}$, the Markov chain reaches upper bound faster than lower bound, or within a same round $r$. So the state space can rewrite as
\begin{equation}\label{S2}\small
\begin{split}
S=
\begin{cases}
\{I_1=1,E_1=k\},\\
\{I_r\in[{2,\lceil\frac{n-1}{k}\rceil-1}],E_r\in[k-r+2,rk-r+1]\},\\
\{I_r\in[\lceil\frac{n-1}{k}\rceil,k+1],E_r\in[k-r+2,n-r]\},\\
\{I_r\in[k+2,n],E_r\in[0,n-r]\}.
\end{cases}
\end{split}
\end{equation}
In this case, $S_{I}$, $S_{II}$, $S_{IV}$ are the same as (\ref{S1type}), while $S_{III}$ rewrites as
\begin{equation}\label{S2type}\small
\begin{split}
S_{III}=
\begin{cases}
\{I_r\in[2,\lceil\frac{n-1}{k}\rceil-2],E_r\in[k-r+2,rk-r+1]\},\\
\{I_r\in[\lceil\frac{n-1}{k}\rceil-1,k],E_r\in[k-r+2,n-r-k]\},\\
\{I_r\in[k+1,n-k-1],E_r\in[1,n-r-k]\}.
\end{cases}
\end{split}
\end{equation}

3) When $\frac{n-2}{2}< k\leq n-3$, we have $\lceil\frac{n-1}{k}\rceil=2$ and thus the Markov chain reaches upper bound in round $2$. The state space can be expressed as (\ref{S2}), while $\{I_r\in[{2,\lceil\frac{n-1}{k}\rceil-1}],E_r\in[k-r+2,rk-r+1]\}$ is an empty set in this case. Accordingly,
\begin{equation}\label{S3type}\small
\begin{split}
&S_{I}=S_{III}=\phi,\\
&S_{II}=\{I_1=1,E_1=k\},\\
&S_{IV}=
\begin{cases}
\{I_r\in[2,k+1],E_r\in[k-r+2,n-r]\},\\
\{I_r\in[k+2,n-1],E_r\in[1,n-r]\},
\end{cases}
\end{split}
\end{equation}

4) When $n-2\leq k\leq n-1$, the Markov chain converges to a single state in each round with state space
\begin{equation}\label{S4}
S=
\begin{cases}
\{I_1=1,E_1=k\},\\
\{I_r\in[2,n],E_r=n-r\}.
\end{cases}
\end{equation}
Under this condition, all the one-step transition probabilities equal to $1$, so there is no need to analyze the subsets of $S$.

\subsection{Failure State in Markov Chain}


At each random selection of $k$ servers, 
there is a probability ${C_{I_r+E_r-2}^{k}\cdot C_{U_r}^{0}}/{C_{n-2}^{k}}$ that the uninformed servers will not be selected in this round. When this event occurs, none of the $k$ servers will request the block using \emph{getdata} message, and thus the number of engaged servers $E_r$ decreases by $1$ until it reaches the lower bound $0$. The failure state $\{I_r,0\}$ ($I_r\in[k+2,n-1]$) in Markov chain describes the situation that block $h$ failed to be propagated to all $n$ servers when the number of engaged servers reduces to $0$. In this situation, all informed servers have already conducted a random selection and sent the \emph{inv} about block $h$, but then they stopped propagating block due to no response from the selected servers.

Blockchain will wait for the generation of the next block to address the propagation failure of block $h$ \cite{1-Antonopoulos}. If the next block is generated by the informed servers, its height will be $h+1$. When a uninformed server of block $h$ receives the \emph{inv} about block $h+1$, it can realize that block $h$ is missed and then request blocks $h$ and $h+1$ using \emph{getdata}. After that, the number of informed servers for block $h$ increases by $1$ and a transition from state $\{I_r,0\}$ to state $\{I_r+1,0\}$ occurs. On the other hand, the next block can also be generated by the uniformed servers, and its height will be $h$. Since we focus on the propagation of block $h$, both of the cases have a similar impact on Markov chain, i.e., state $\{I_r,0\}$ will transit to $\{I_r+1,0\}$ either when block $h$ or block $h+1$ is introduced to a uninformed server.

1) \emph{Transition probability of failure state}. The Markov chain for the next block propagation is the same as that for the first block propagation, and it will affect the failure state of the first one. By the end of the next block propagation, the Markov chain for the first block can move from state $\{I_r,0\}$ to $\{I_r+i,0\}$ ($i\in[0,n-I_r]$), where $i$ is the number of servers that receives block $h$ in the next block propagation. Accordingly, the transition probabilities of failure states are
\begin{equation}\label{second}
\begin{split}
P\{I_r+i,0\mid I_r,0\}=\sum\limits_{j=\max\{k+2,i\}}^{I_r+i}p_{j,0}\frac{{C_{I_r}^{j-i} C_{n-I_r}^{i}}}{{C_{n}^{j}}},
\end{split}
\end{equation}
where $I_r\in[k+2,n-1]$, $i\in[0,n-I_r]$. $p_{j,0}=P\{j,0\mid 1,k\}$ denotes the $(j-1)$-step transition probability that can be calculated by (\ref{onestep}). Specifically, equation (\ref{second}) represents that the next block has reached a total of $j$ servers, which consists of $j-i$ servers receiving block $h$ in the first propagation and $i$ servers receiving block $h$ in the next propagation. Taking $n=10$, $k=2$, $I_r=6$, $i=2$ as an example, we can specify (\ref{second}) as $P\{8,0\mid6,0\}=\sum\limits_{j=4}^{8}p_{j,0}\frac{{C_{6}^{j-2} C_{4}^{2}}}{{C_{10}^{j}}}$. It means that there are four uninformed servers after the first propagation, and then two of them become informed servers after the next propagation.

\begin{figure}[t]
\setlength{\abovecaptionskip}{0.cm}
\setlength{\belowcaptionskip}{-0.cm}
\captionsetup{font={footnotesize}}
\begin{center}
 \includegraphics[width=9cm]{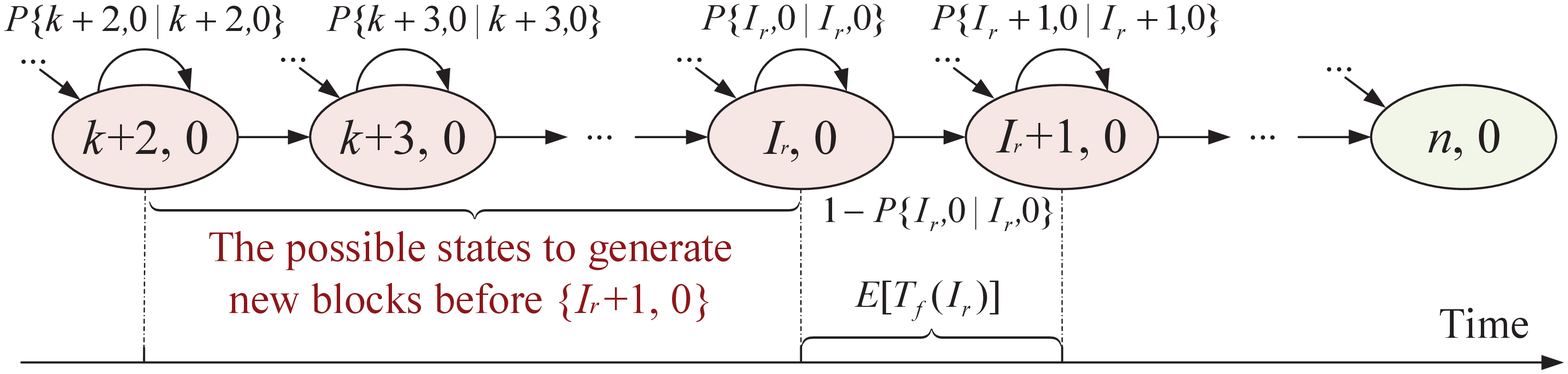}
 \end{center}
 \caption{{The transition diagram of failure states.}}
\label{failure}
\end{figure}

2) \emph{Expected time to go from state $\{I_r,0\}$ to state $\{I_r+1,0\}$}. 
As shown in Fig. \ref{failure}, before going from state $\{I_r,0\}$ to state $\{I_r+1,0\}$, the first time that the Markov chain enters a failure state could be in any of states $\{k+2,0\}$, $\{k+3,0\}$, ..., $\{I_r,0\}$ with probability
\begin{equation}\label{first}
\begin{split}
p_{l,0}^{*}=P^{*}\{l,0\mid 1,k\},~l\in[k+2,I_r],
\end{split}
\end{equation}
where {\small$\sum\limits_{l=k+2}^{I_r}P^{*}\{l,0\mid 1,k\}=1$} denotes the normalized $(l-1)$-step transition probabilities, which are derived by calculating the one-step probabilities using the condition that the Markov chain must pass through state $\{I_r,0\}$. For example, if we know that the Markov chain in Fig. \ref{Markovfig} must pass through state $\{6,0\}$, then the previous states of $\{6,0\}$ should satisfy $I_r+E_r\leq6$, shown in the shaded area of Fig. \ref{Markovfig}. Solving $I_r+E_r\leq6$ and $I_r=r$ yields $E_r\leq 6-r$, so the one-step transition probabilities can be recalculated based on $I_r\in[2,6]$, $E_r\in[\max\{k-r+2, 0\},\min\{rk-r+1,{6-r}\}]$. Then, we can obtain {\small$\sum\limits_{l=4}^{6}P^{*}\{l,0\mid 1,2\}=1$}.

After entering a failure state, the Markov chain must wait for the generation of new blocks until it can transit to state $\{I_r+1,0\}$. Due to the self-loop probability in failure state, the possible number of blocks that is generated until $\{I_r+1,0\}$ occurs belongs to $[1,+\infty)$, and the last block will result in the transition from state $\{I_r,0\}$ to state $\{I_r+1,0\}$.
The probability that the last block is generated in state $\{L,0\}$ is
\begin{equation}\label{event}
\begin{split}
P\left\{ {L=l} \right\}=\frac{\alpha(l)\beta(l,I_r)}{\beta(l,l)},~l\in[k+2,I_r],
\end{split}
\end{equation}
where $\sum\limits_{l=k+2}^{I_r}P\left\{ {L=l} \right\}=1$ with
\begin{equation}\label{alpha}
\begin{split}
&\alpha(l)=
\begin{cases}
p_{k+2,0}^{*},~~~~~~~~~~~~~~~~~~~~~~~~~~~~\text{if}~l=k+2,\\
\sum\limits_{i=k+2}^{l-1}\alpha(i)\frac{P\{l,0\mid i,0\}}{1-P\{i,0\mid i,0\}}+p_{l,0}^{*},~~\text{if}~l\in[k+3,I_r],
\end{cases}\\
&\beta(l,I_r)=1-\sum\limits_{j=l}^{I_r}P\{j,0\mid l,0\},~\beta(l,l)=1-P\{l,0\mid l,0\}.
\end{split}
\end{equation}
Specifically, $\alpha(l)$ is the probability that a block is generated in state $\{l,0\}$, without knowing that whether this block will result in the transition from $\{I_r,0\}$ to $\{I_r+1,0\}$. As a complement to $\alpha(l)$, $\beta(l,I_r)/\beta(l,l)$ is the probability that $\{l,0\}$ transits to a state greater than $\{I_r,0\}$, after going through $0,1,2, \cdots ,\infty$ self-loops. Note that $i< l\leq j$ in (\ref{alpha}), since $\{i,0\}$ and $\{j,0\}$ denote the past and future states of $\{l,0\}$ respectively. Taking $n=10$, $k=2$, $I_r=6$, $L=5$ as an example, we have
\begin{equation}\label{eventexample}\small
\begin{split}
&P\{L=5\}=\\
&\left(p_{4,0}^{*}\frac{P\{5,0\mid 4,0\}}{1-P\{4,0\mid 4,0\}}+p_{5,0}^{*}\right)\frac{1-\sum\limits_{j=5}^{6}P\{j,0\mid 5,0\}}{1-P\{5,0\mid 5,0\}}.
\end{split}
\end{equation}
The expression in the parentheses of (\ref{eventexample}) is the probability that the block is generated in state $\{5,0\}$. The expression outside the parentheses is the probability that $\{5,0\}$ transits to a state between $\{7,0\}$ and $\{10,0\}$ by the end of this block propagation, and thus the transition from $\{6,0\}$ to $\{7,0\}$ must have occurred during this propagation.

Now we should determine the transition from state $\{I_r,0\}$ to state $\{I_r+1,0\}$ occurs in which round of the block propagation. Under the condition that the block for $\{I_r+1,0\}$ is generated in state $\{L,0\}$, the probability that the transition from $\{I_r,0\}$ to $\{I_r+1,0\}$ occurs in round $R$ is
\begin{equation}\label{PR}
\begin{split}
~~~&P\{R=r\mid L=l\}=\\
~~~&\begin{cases}
0,~~~~~~~~~~~~~~~~~~~~~~~~~~~~~~~\text{if}~r\in[1,I_r-l],\\
\frac{C_{r-1}^{I_r-l} A_{l}^{r-(I_r-l+1)} A_{n-l}^{I_r-l+1}}{A_{n}^{r}\beta(l,I_r)},~~~~\text{if}~r\in[I_r-l+1,k+2],\\
\left(1-\sum\limits_{j=k+2}^{r-1}p_{j,0}\right)\frac{C_{r-1}^{I_r-l} A_{l}^{r-(I_r-l+1)} A_{n-l}^{I_r-l+1}}{A_{n}^{r}\beta(l,I_r)},\\
~~~~~~~~~~~~~~~~~~~~~~~~~~~~~~~~~~\text{if}~r\in[k+3,I_r+1],
\end{cases}
\end{split}
\end{equation}
where $I_r\in[k+2,k+l+1]$ and $l\in[k+2,I_r]$. $A_{n}^{r}$ denotes the number of ordered arrangements that selecting $r$ servers from $n$ servers at random, known as permutation. Note that when $I_r\in[k+l+2,n-1]$, we have $I_r-l+1>k+2$ in (\ref{PR}). In this case, $r\in[I_r-l+1,k+2]$ is an empty set, and $r\in[k+3,I_r+1]$ changes to $r\in[I_r-l+1,I_r+1]$. We also use $n=10$, $k=2$, $I_r=6$, $L=5$ as an example to specify (\ref{PR}) as follows:
\begin{equation}\label{PRexample}\small
\begin{split}
&P\{R=1\mid L=5\}=0,~P\{R=2\mid L=5\}=\frac{A_{5}^{0}A_{5}^{2}}{A_{10}^{2}\beta(5,6)},\\
&...,~P\{R=7\mid L=5\}=\left(1-\sum\limits_{j=4}^{6}p_{j,0}\right)\frac{C_{6}^{1}A_{5}^{5}A_{5}^{2}}{A_{10}^{7}\beta(5,6)},
\end{split}
\end{equation}
where $\beta(5,6)$ is the probability that state $\{5,0\}$ transits to state $\{7,0\}$ during a block propagation, so this propagation must have reached two uniformed servers. We have $P\{R=1\mid L=5\}=0$ since one round cannot reach two servers. $P\{R=2\mid L=5\}$ means that two rounds reach two uninformed servers successively. $P\{R=7\mid L=5\}$ means that seven rounds reach five informed servers and two uninformed servers, where one of the uninformed servers must be reached in round $7$ and the other one can be reached in any of the six rounds with $C_{6}^{1}$.

Based on (\ref{event}) and (\ref{PR}), the expected time to go from failure state $\{I_r,0\}$ to state $\{I_r+1,0\}$ is given by
\begin{equation}\label{T0}
\begin{split}
&E[T_f(I_r)]=\sum\limits_{l=k+2}^{I_r}P\{L=l\}\times\sum\limits_{r=I_r-l+1}^{I_r+1}P\{R=r\mid L=l\}\\
&\times\sum\limits_{j=I_r-l}^{r-1}\frac{C_{j}^{I_r-l}}{C_{r-1}^{I_r-l}}E[T_j\mid R=r]+\frac{P\{L=I_r\}P\{I_r,0\mid I_r,0\}}{1-P\{I_r,0\mid I_r,0\}}t_b,
\end{split}
\end{equation}
where $\frac{P\{L=I_r\}P\{I_r,0\mid I_r,0\}}{1-P\{I_r,0\mid I_r,0\}}t_b$ is the time incurred by the self-loop of $\{I_r,0\}$, and each self-loop has an additional block generation time $t_b$. $E[T_j\mid R=r]$ is the expected time from state $\{I_r,E_r\}$ to state $\{I_{r+1},E_{r+1}\}$, conditional on the transition from $\{I_r,0\}$ to $\{I_r+1,0\}$ happens in the round $r$ of a block propagation. Note that when we know that a block propagation has reached round $r$, the failure states must not have occurred before round $r$. Therefore, to obtain $E[T_j\mid R=r]$, we should first normalize the one-step probabilities of states $\{I_r,1\}$ ($I_r\in[k+2,n-2]$) based on the condition $P\{I_r+1,0\mid I_r,1\}=0$, and then substituting the one-step probabilities into (\ref{ETr}).

\section{Performance Analysis of Blockchain}

\subsection{Block Propagation Performance}

1) \emph{The increasing rate of informed servers}. The increasing rate is the number of informed servers increased per second. For a given round $r$, the increasing rate is $1/E[T_r]$, in which $E[T_r]$ is the expected time to go from state $\{I_r,E_r\}$ to state $\{I_{r+1},E_{r+1}\}$ that can be expressed as
\begin{equation}\label{ETr}
\begin{split}
E[T_r]=
\begin{cases}
s_b/(\lambda_d E_1),~~~~~~~~~~~~~~~~~~~~~~~~~\text{if}~I_r=1,\\
\sum\limits_{E_r = f_{\max}(r)}^{f_{\min}(r)}P\{ {I_r,E_r\mid I_1,E_1}\}\cdot s_b/(\lambda_d E_r),\\
~~~~~~~~~~~~~~~~~~~~~~~~~~~~~~~~~~~~~~\text{if}~I_r\in[2,k+1],\\
\sum\limits_{E_r = 1}^{f_{\min}(r)}P\{ {I_r,E_r\mid I_1,E_1}\}\cdot s_b/(\lambda_d E_r)\\
+P\{ {I_r,0\mid I_1,E_1}\}\cdot E[T_f(I_r)],~\text{if}~I_r\in[k+2,n),
\end{cases}
\end{split}
\end{equation}
where $P\{ {I_r,E_r\mid I_1,E_1}\}$ is the ($r-1$)-step transition probability that can be obtained by raising one-step transition probability matrix to the power $r-1$. To reflect the dynamic change of informed servers over time, the cumulative distribution function of time is derived as follows, namely \emph{the number of informed servers vs. time}:
\begin{equation}\label{cumulative}
\begin{split}
I_r(t)=
\begin{cases}
1,~~~t=0,\\
i,~~~t=\sum\limits_{r=1}^{i-1}E[T_r],
\end{cases}
\end{split}
\end{equation}
where $i\in[2,n]$.

2) \emph{Block propagation delay and failure probability}. Block propagation delay $t_p$ is defined as the time from the generation of block $h$ until it have been propagated to all $n$ servers. Based on (\ref{cumulative}), it is straightforward to give $t_p$ by
\begin{equation}\label{Tp}
\begin{split}
t_p=\sum\limits_{r=1}^{n-1}E[T_r].
\end{split}
\end{equation}
Now we analyze the lower bound of $t_p$. According to (\ref{ETr}), $E[T_r]$ is affected by the number of engaged servers $E_r$, where $E_r\in[k-r+2, rk-r+1]\cap[0, n-r]$ in round $r$. Therefore, $t_p$ can be considered as a function of $k$, and we rewrite it as $t_p(k)$ ($k\in[1,n-1]$). Based on $E_r\in[k-r+2, rk-r+1]\cap[0, n-r]$, we can know that $E_r$ increases with $k$ before reaching $n-r$. Meanwhile, (\ref{ETr}) shows that $E[T_r]$ decreases with $E_r$. So in summary, $t_p(k)$ decreases monotonically in the range of $k\in[1,n-1]$, and thus the lower bound of propagation delay is $t_p(n-1)$. Considering the complexity of the relationship between $t_p$ and $k$, we verified the monotonicity of $t_p(k)$ using numerical calculation, shown in Fig. \ref{propagation} (d). To derive the expression of $t_p(n-1)$, we recall that the Markov chain converges to a single state in each round when $k=n-1$, with state space $\{1,n-1\}$, $\{2,n-2\}$, ...., $\{n,0\}$. Then, using equation $E[T_r]=s_b/(\lambda_d E_r)$ to calculate the expected time from round $r$ to $r-1$, we can obtain
\begin{equation}\label{Tpbound}
\begin{split}
t_p(n-1)=\frac{s_b}{\lambda_d}\sum\limits_{E_r=1}^{n-1}\frac{1}{E_r}\approx\frac{s_b}{\lambda_d}\left[\ln(n-1)+\frac{1}
{2(n-1)}+\gamma\right],
\end{split}
\end{equation}
where $\sum\limits_{E_r=1}^{n-1}\frac{1}{E_r}$ is the harmonic series that is approximated by a logarithmic function with Euler's constant $\gamma\approx0.5772$ \cite{1-Havil}.

Propagation failure probability $p_f$ is defined as the probability that block $h$ failed to be propagated to all $n$ servers when the number of engaged servers reduces to $0$. To determine $p_f$, we can first calculate the complementary probability that the Markov chain has reached state $\{n,0\}$ successfully by the end of a block propagation, which is the $(n-1)$-step probability $P\{n,0\mid 1,k\}$. Then, propagation failure probability is
\begin{equation}\label{pf}
\begin{split}
p_f=1-P\{n,0\mid 1,k\}.
\end{split}
\end{equation}

3) \emph{Forking probability}. For a given round of block propagation, there are $r$ servers that have received block $h$, so they mine the next block based on block $h$. On the other hand, there are $n-r$ servers that do not received block $h$, and they mine the next block based on block $h-1$. If the next block is generated based on block $h-1$, a forking problem occurs.
Let $T_k(r)$ denote the time to generate a fork in round $r$. According to \cite{1-Nakamoto}, \cite{7-yixinli}, $T_k(r)$ is exponentially distributed, and its expectation is inversely proportional to block generation rate. Let $\lambda_b$ be the total block generation rate of a consensus domain, where $\lambda_b=1/t_b$. During the propagation of block $h$, $\lambda_b$ is split into two parts, in which $\frac{r}{n}\lambda_b$ tries to extend block $h$ and $\frac{n-r}{n}\lambda_b$ tries to extend block $h-1$. The time to generate a fork is affected by the block generation rate on block $h-1$, so we can obtain $E[T_k(r)]=1/\left(\frac{n-r}{n}\lambda_b\right)$. Note that $\frac{n-r}{n}\lambda_b$ will change with round $r$, and thus each round has different forking probability. Using the complementary event that the fork does not occur in all rounds, forking probability is given by
\begin{equation}\label{pk}
\begin{split}
p_k=&1-P\{T_k(1)>t_1,T_k(2)>t_2,...,T_k(n-1)>t_{n-1}\}\\
=&1-P\{T_k(1)>t_1\}\times P\{T_k(2)>t_2\mid T_k(1)>t_1\}\times...\times\\
&P\{T_k(n-1)>t_{n-1}\mid T_k(1)>t_1,...,T_k(n-2)>t_{n-2}\}\\
=&1-\prod\limits_{r=1}^{n-1}P\{T_k(r)>t_r\}~\text{(by memoryless property)}\\
=&1-\exp\left(-\sum\limits_{r=1}^{n-1}\frac{n-r}{n}\lambda_b t_r\right),
\end{split}
\end{equation}
where $t_r=E[T_r]$ denotes the expected time from round $r$ to round $r+1$. $\{T_k(r)>t_r\}$ is an event that the time to generate a fork is larger than $t_r$, so the fork does not occur in round $r$. Meanwhile, the expected number of forks that is generated during the propagation of block $h$ is
\begin{equation}\label{nk}
\begin{split}
n_k=\sum\limits_{r=1}^{n-1}\frac{n-r}{n}\lambda_b t_r.
\end{split}
\end{equation}

\subsection{Transaction-Processing Capability}

1) \emph{Transaction throughput}. Transaction throughput $\theta$ is the maximum number of transactions that can be processed by blockchain per second, known as transaction per second (TPS). To calculate $\theta$, we can multiply the number of transactions in a block by the number of valid blocks generated per second. The number of transactions in a block can be obtained by $n_t=(s_{b}-s_{h})/s_{t}$, where $s_b$ denotes the size of a block, $s_{h}$ denotes the size of block header, and $s_{t}$ denotes the size of a transaction. On the other hand, the number of valid blocks generated per second is affected by block generation rate. We know that all servers constantly mine new blocks with block generation rate $\lambda_b$, which means that there are a total of $\lambda_b$ blocks generated per second. Among them, forks are invalid blocks that cannot confirm transactions. By means of (\ref{nk}), the ratio of valid blocks to invalid blocks (forks) is $1:n_k$. Therefore, the number of valid blocks generated per second is $\lambda_b/(1+n_k)$. Based on the analysis, transaction throughput is given by
\begin{equation}\label{theta}
\begin{split}
\theta=\frac{\lambda_bn_t}{1+n_k}.
\end{split}
\end{equation}
Substituting (\ref{nk}) into (\ref{theta}) and let block generation rate $\lambda_b \to \infty$, the upper bound of transaction throughput is derived as follows:
\begin{equation}\label{thetaupper}
\begin{split}
\overline{\theta}&=\lim_{\lambda_b \to \infty }\frac{\lambda_bn_t}{1+\sum\limits_{r=1}^{n-1}\frac{n-r}{n}\lambda_b t_r}=\frac{n_t}{\sum\limits_{r=1}^{n-1}\frac{n-r}{n} t_r}.
\end{split}
\end{equation}
In the numerator, $n_t\!=({s_{b}\!-\!s_{h}})/{s_{t}}$. In the denominator, $t_r=E[T_r]$ is the expected time from round $r$ to round $r+1$, and its value mainly depends on $s_b/(\lambda_d E_r)$ in (\ref{ETr}), where $E_r\in[k-r+2, rk-r+1]\cap[0, n-r]$. Since both numerator and denominator have $s_{b}$, and $s_{b}\gg s_{h}$, the impact of $s_{b}$ on $\overline{\theta}$ has been counteracted. Therefore, $\overline{\theta}$ is determined by the total number of servers $n$, the number of selected servers $k$, network data rate $\lambda_d$, and transaction size $s_{t}$.

2) \emph{Confirmation delay}. In blockchain, a transaction has one confirmation once it is included in a valid block. Then, with the accumulation of other valid blocks, the probability of malicious modification will decrease exponentially \cite{1-Nakamoto}. When the number of confirmations reaches a given threshold $m$, the transaction is considered to be irreversible. Accordingly, confirmation delay $t_c$ is defined as the time from a transaction is broadcast to the network until it has $m$ confirmations. Based on the definition, $t_c$ is equal to the time that a transaction is processed by servers plus the time to wait for $m$ valid blocks, which is given by
\begin{equation}\label{tc}
\begin{split}
t_c=t_w+\frac{1+n_k}{\lambda_b}m,
\end{split}
\end{equation}
where $t_w$ denotes the transaction waiting time in transaction pool, which is influenced by transaction fee \cite{1-Antonopoulos}. $(1+n_k)/\lambda_b$ is the time to generate a valid block. By means of (\ref{thetaupper}), the lower bound of confirmation delay is
\begin{equation}\label{tclower}
\begin{split}
\underline{t_c}=t_w+{\sum\limits_{r=1}^{n-1}\frac{n-r}{n} t_r}\cdot m.
\end{split}
\end{equation}


\subsection{Blockchain Security Analysis}

1) \emph{The probability of malicious modification}. A typical malicious modification of blockchain ledger is the double-spending attack in Bitcoin, which inserts a pair of conflicting transactions into two chains in parallel through deliberate forking. When the first transaction is confirmed, the merchant will deliver products or services to the payer (attacker). At this time, the attacker can broadcast a longer chain that contains a conflicting transaction to replace the original main chain, so that the first transaction becomes invalid.

\begin{figure}[t]
\setlength{\abovecaptionskip}{0.cm}
\setlength{\belowcaptionskip}{-0.2cm}
\captionsetup{font={footnotesize}}
\begin{center}
 \includegraphics[width=9.2cm]{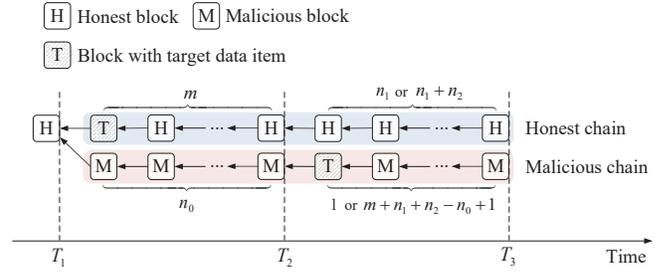}
 \end{center}
 \caption{{The malicious modification of data ownership through forking.}}
\label{copyright}
\end{figure}

Through deliberate forking, the data items stored in blockchain may become the target of malicious servers to modify the ownership, shown in Fig. \ref{copyright}. The main events in attack process are:
(i) At time $T_{1}$, the malicious server has a target data item for ownership modification, so it begins to build an offline malicious chain to compete with the honest chain.
(ii) At time $T_{2}$, the target data item reaches confirmation threshold $m$, so the data producer decrypts the data item. After that, the malicious server can add the target data item with a modified owner ID into the malicious chain and keep on generating new block to outpace the honest chain.
(iii) At time $T_{3}$, the malicious chain outpaces honest chain by one block, so the malicious server broadcasts it to the other servers. Since all servers follow the longest chain rule, the target data item in the honest chain becomes invalid.

Let $\lambda_h$ and $\lambda_m$ denote the block generation rate of honest and malicious servers respectively, where $\lambda_h+\lambda_m=\lambda_b$. Based on our previous analysis for double-spending \cite{7-yixinli}, the competition between honest and malicious chains can be modeled as independent Bernoulli trials, where a honest block occurs with probability $p=\lambda_h/(\lambda_h+\lambda_m)$ and a malicious block occurs with probability $q=\lambda_m/(\lambda_h+\lambda_m)$. In this work, we extend this model to analyze the probability of malicious modification on data ownership while considering the impacts of: (i) The forking problem in honest chain due to propagation delay. (ii) The unavailability of data item before having $m$ confirmations. For the first problem, recalling that the ratio of valid blocks to invalid blocks (forks) is $1:n_k$ based on (\ref{nk}). Therefore, the number of valid blocks generated by honest servers per second is $\lambda_h'=\lambda_h/(1+n_k)$. Under the impact of forking, $p$ and $q$ in Bernoulli trials rewrite as
\begin{equation}\label{pq}
\begin{split}
p=\frac{\lambda_h'}{\lambda_h'+\lambda_m}=\frac{\lambda_h}{\lambda_h+(1+n_k)\lambda_m},\\
q=\frac{\lambda_m}{\lambda_h'+\lambda_m}=\frac{(1+n_k)\lambda_m}{\lambda_h+(1+n_k)\lambda_m}.
\end{split}
\end{equation}

For the second problem, we know that the data item can be encrypted by data producer for ownership protection until it has $m$ confirmations. In this case, the malicious server cannot broadcast its chain before $T_2$, since it has not yet obtained the data item. Let $N_0$ be the number of malicious blocks until the $m$th honest block occurs, which is a negative binomial random variable with probability mass function
\begin{equation}\label{negative}
P\{ N_0=n_0\}=C_{m+n_0-1}^{m-1}{p^m}{q^{n_0}},~n_0=0,1,2,...,\infty,
\end{equation}
where $C_{m+n_0-1}^{m-1}$ denotes the number of combinations that selecting $m-1$ blocks from $m+n_0-1$ blocks at random. This means that the honest and malicious chains have $m$ and $N_0$ blocks respectively from $T_1$ to $T_2$. If $N_0>m$, the malicious server should generate one block at least after $T_2$ to include the data item; if $N_0\leq m$, the malicious server should catch up the difference of $m-N_0+1$ blocks between two chains, referring to the gambler's ruin problem \cite{1-Ross}. The probability to catch up the difference of $D$ blocks is
\begin{equation}\label{gambler}
P\{D=d\}=
\begin{cases}
(q/p)^d,~~~~~~~~\text{if}~p>q,\\
1,~~~~~~~~~~~~~~\,\text{if}~p\leq q.
\end{cases}
\end{equation}
Since the success probability is equal to $1$ when $p\leq q$, we can only analyze the case when $p>q$. Using (\ref{negative}) and (\ref{gambler}), the probability of malicious modification is
\begin{equation}\label{pm}
\begin{split}
p_m=&P\{N_0>m\}p(n_0)+P\{N_0\leq m\}P\{D=m-N_0+1\} \\
= &\sum\limits_{n_0={m+1}}^\infty C_{m+n_0-1}^{m-1}{p^m}{q^{n_0}}p(n_0)+\sum\limits_{n_0=0}^m C_{m+n_0-1}^{m-1}{p^{n_0-1}}{q^{m+1}},
\end{split}
\end{equation}
where $p(n_0)$ is the probability that the malicious chain wins after including the data item, and it can be expressed as
\begin{equation}\label{pn0}
\begin{split}
p(n_0)=&P\{N_1\leq n_0-m\}\\
&+P\{N_1>n_0-m\}P\{D=m+N_1-n_0+1\} \\
=&\sum\limits_{n_1=0}^{n_0-m} {p^{n_1}}q+\sum\limits_{n_1=n_0-m+1}^\infty {p^{n_1}}q\left(\frac{q}{p}\right)^{m+n_1-n_0+1}\\
=&\frac{1-p^{n_0-m+1}}{1-p}q+p^{n_0-m-1}\frac{q^3}{1-q}\\
=&1-p^{n_0-m-2}\left(p^3-q^3\right).
\end{split}
\end{equation}
Note that $N_1$ denotes the number of honest blocks until one malicious block occurs after $T_2$, and its probability mass function is given by
\begin{equation}\label{negativen1}
P\{ N_1=n_1\}={p^{n_1}}q,~n_1=0,1,2,...,\infty.
\end{equation}

As a comparison of ($\ref{pm}$), the probability of malicious modification without data encryption is
\begin{equation}\label{pw}
p_w=P\{D=1\}=q/p.
\end{equation}
Without the encryption from data producer, the malicious server can obtain the data item at time $T_1$, so it only needs to outpace honest chain by one block.


2) \emph{Fault tolerance}. The fault tolerance of consensus mechanism is defined as the lowest consensus resources needed by malicious server to guarantee the success of data modification \cite{1-Antonopoulos}. The ideal fault tolerance of proof-type consensus mechanism is $50\%$ of the total resources, which indicates a threshold that the malicious server can modify arbitrary data items with probability $1$, known as a $50\%$ attack. Under the impact of forking, the fault tolerance will be lower than $50\%$, since the resources spent on forks are invalid. To calculate fault tolerance, we notice that the probability of malicious modification is just equal to $1$ when $p=q$ in (\ref{gambler}). Letting $p=q$ in (\ref{pq}), we can obtain $\lambda_h=(1+n_k)\lambda_m$. Then, using the fact that block generation rate is proportional to consensus resource, fault tolerance is given by
\begin{equation}\label{tolerance}
\text{FT}=\frac{\lambda_m}{\lambda_h+\lambda_m}=\frac{\lambda_m}{(1+n_k)\lambda_m+\lambda_m}=\frac{1}{2+n_k},
\end{equation}
which satisfies $\text{FT}=0.5$ when fork ratio $n_k=0$.

\section{Numerical Results and Discussions}

In this section, we use Matlab to calculate the closed-form expressions of performance metrics for demonstrating the performance bounds and trade-offs of blockchain.

\subsection{Parameter Settings and Initialization}

We evaluate the performance of blockchain in a consensus domain with $n$ edge servers, which follow the proof-type consensus mechanism and the gossiping-based block propagation. The parameter settings are based on literatures \cite{1-Antonopoulos} and \cite{4-JinkeRen}, listed in Table I. For initialization, one of the important procedures is to generate the one-step transition probability matrix of Markov chain in Matlab using equation (\ref{onestep}), where the four types of states can be determined using equations (\ref{S1})-(\ref{S4}). Another important procedure is to calculate the expected time to leave a failure state using equations (\ref{event}), (\ref{PR}), and (\ref{T0}).



\subsection{Propagation Performance Evaluations}

\begin{figure*}[t]
\setlength{\abovecaptionskip}{0.cm}
\setlength{\belowcaptionskip}{-0cm}
\captionsetup{font={footnotesize}}
\centering
\subfigure[Informed servers vs. time]{
\includegraphics[width=4.1cm]{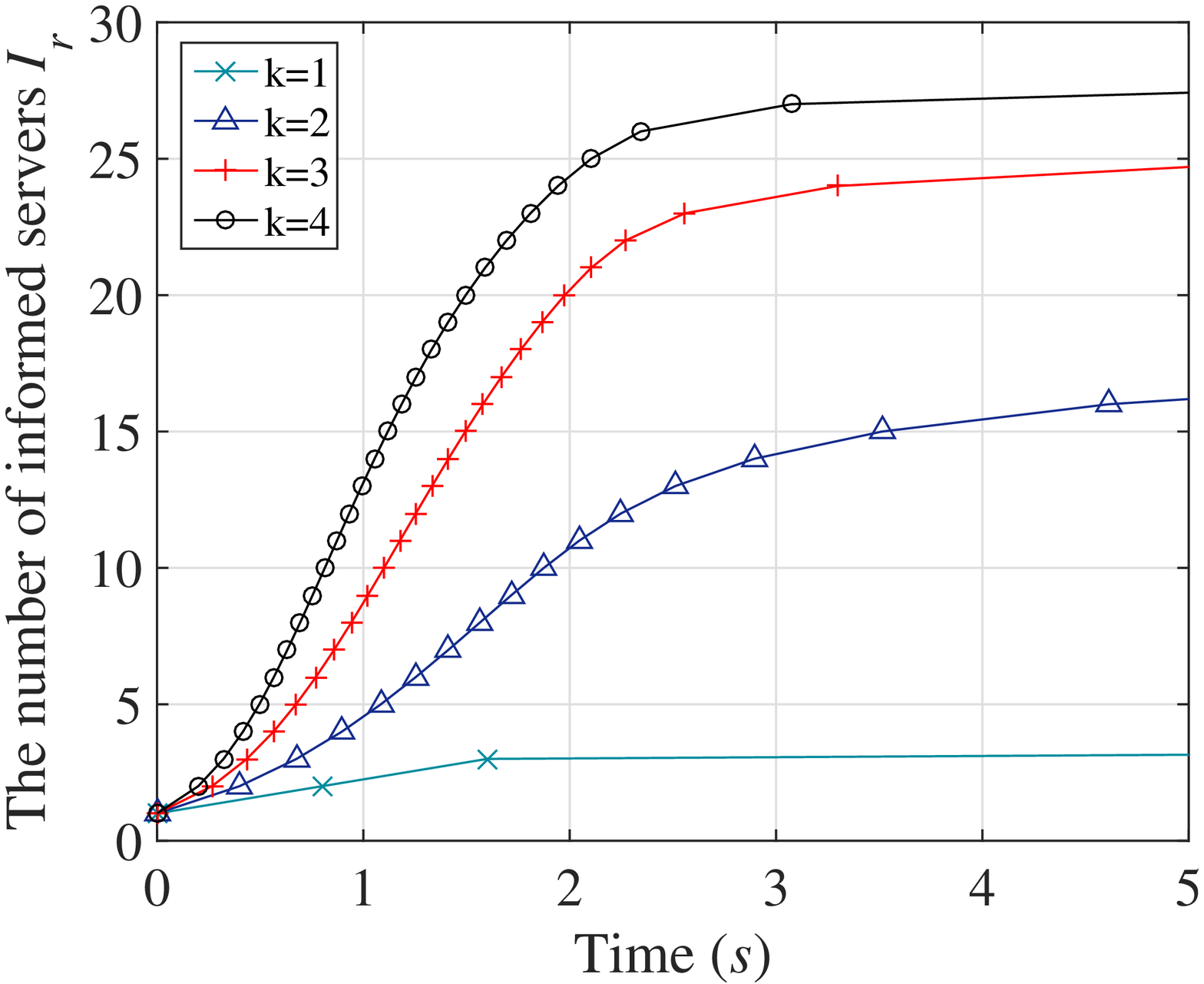}
}
\subfigure[Informed servers vs. time]{
\includegraphics[width=4.1cm]{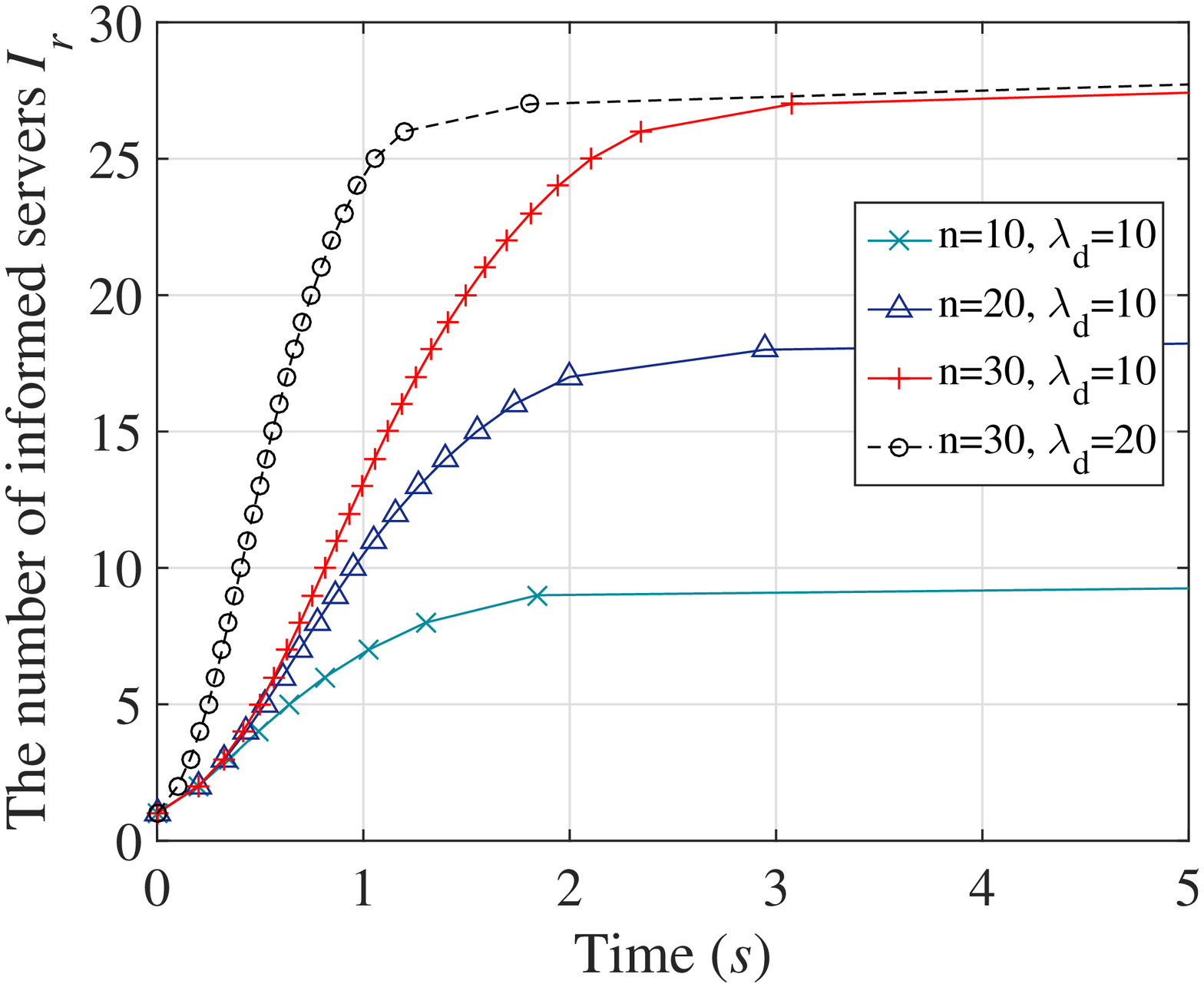}
}
\subfigure[Propagation delay vs. $n$]{
\includegraphics[width=4.1cm]{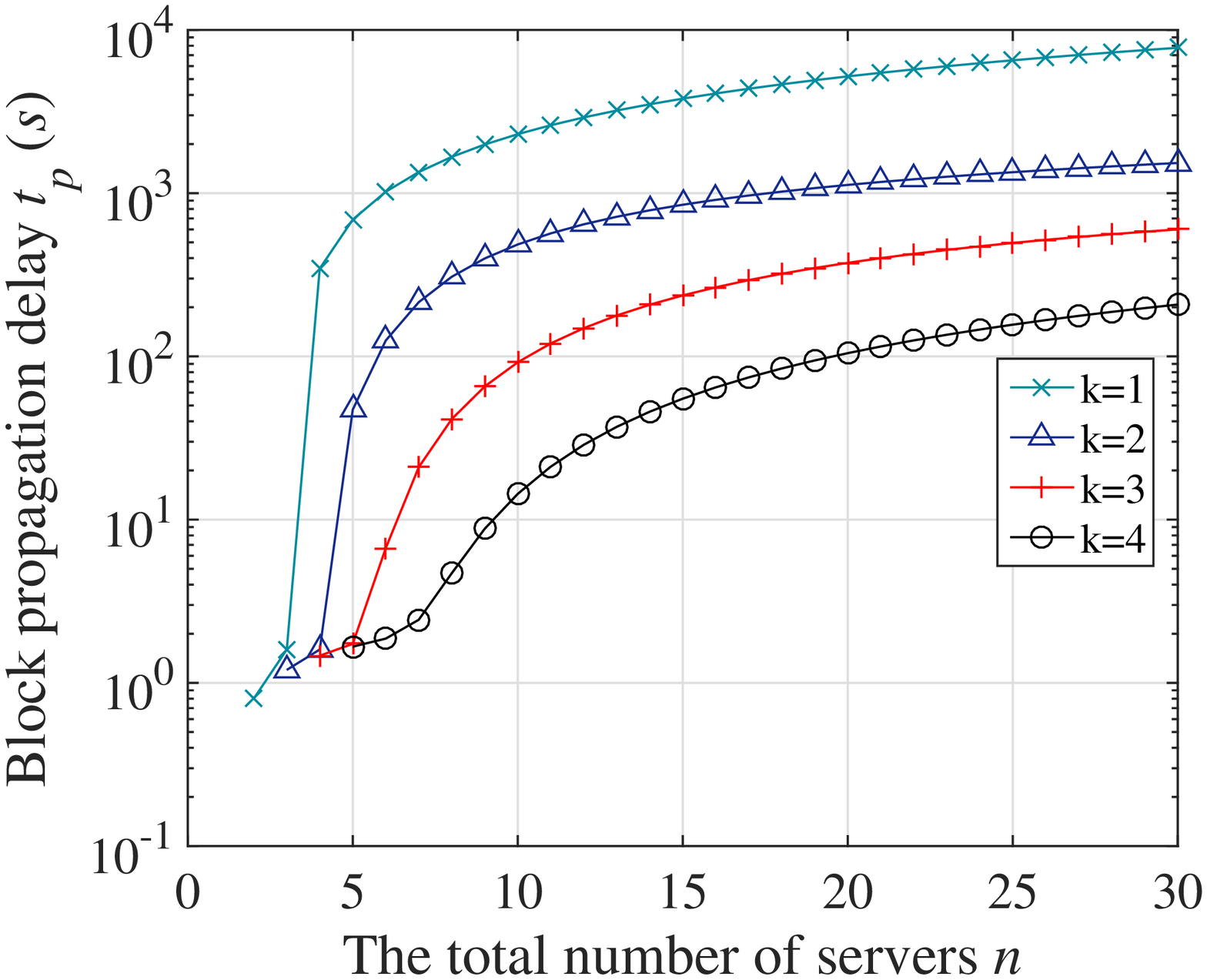}
}
\subfigure[Propagation delay vs. $k$]{
\includegraphics[width=4.1cm]{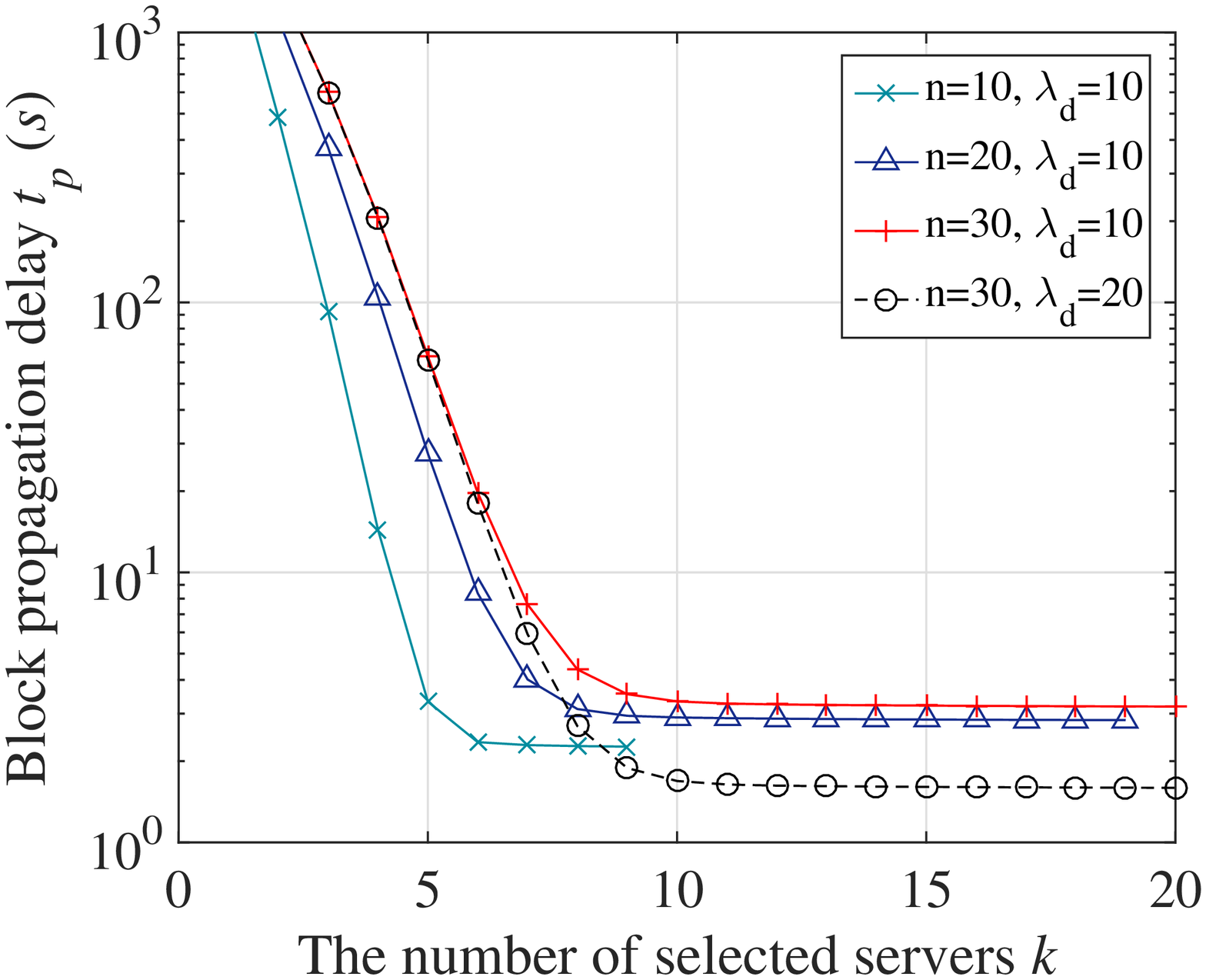}
}\\
\subfigure[Failure probability vs. $n$]{
\includegraphics[width=4.1cm]{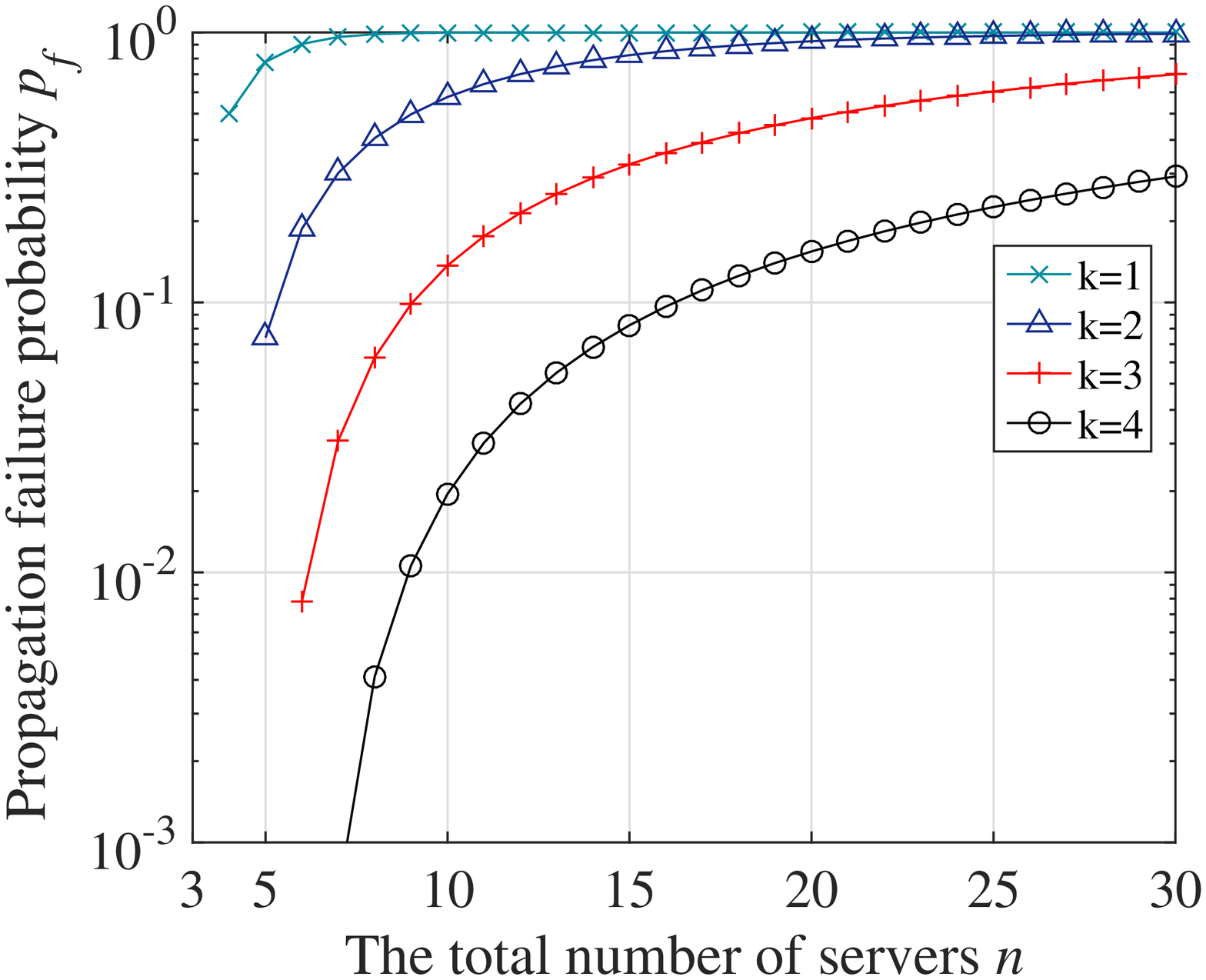}
}
\subfigure[Failure probability vs. $k$]{
\includegraphics[width=4.1cm]{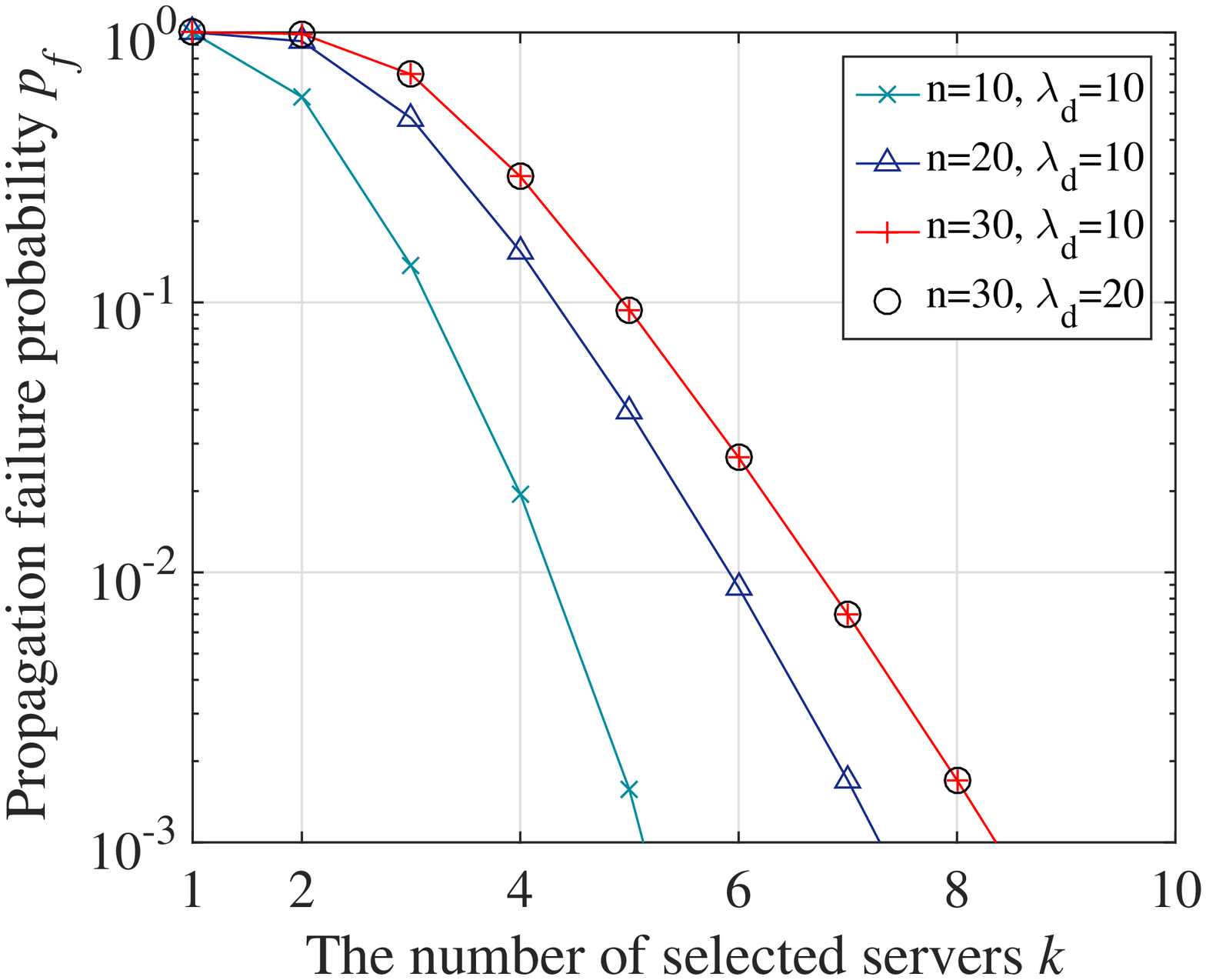}
}
\subfigure[Minimum $k$ to satisfy accuracy]{
\includegraphics[width=4.1cm]{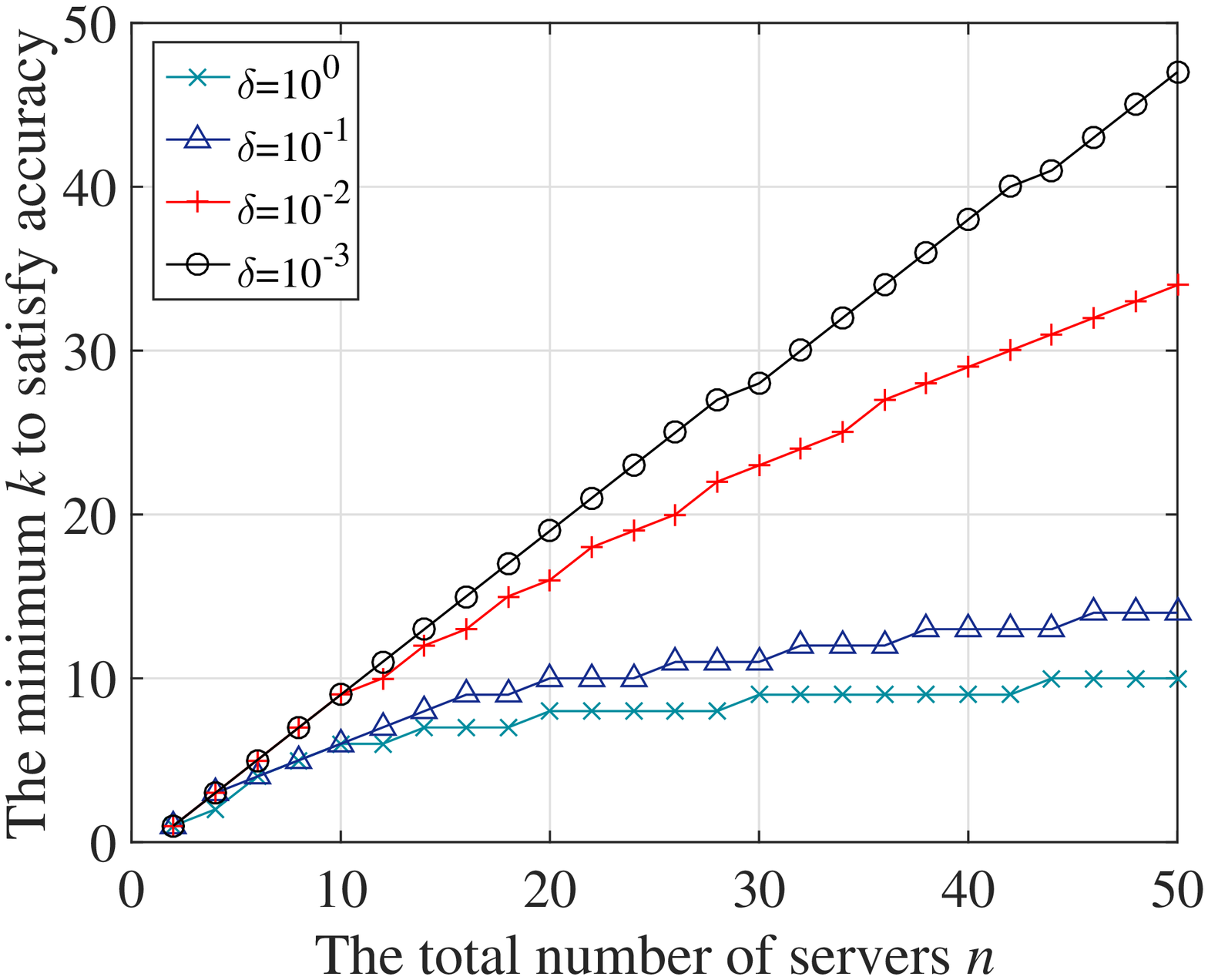}
}
\subfigure[Delay lower bound vs. $n$]{
\includegraphics[width=4.1cm]{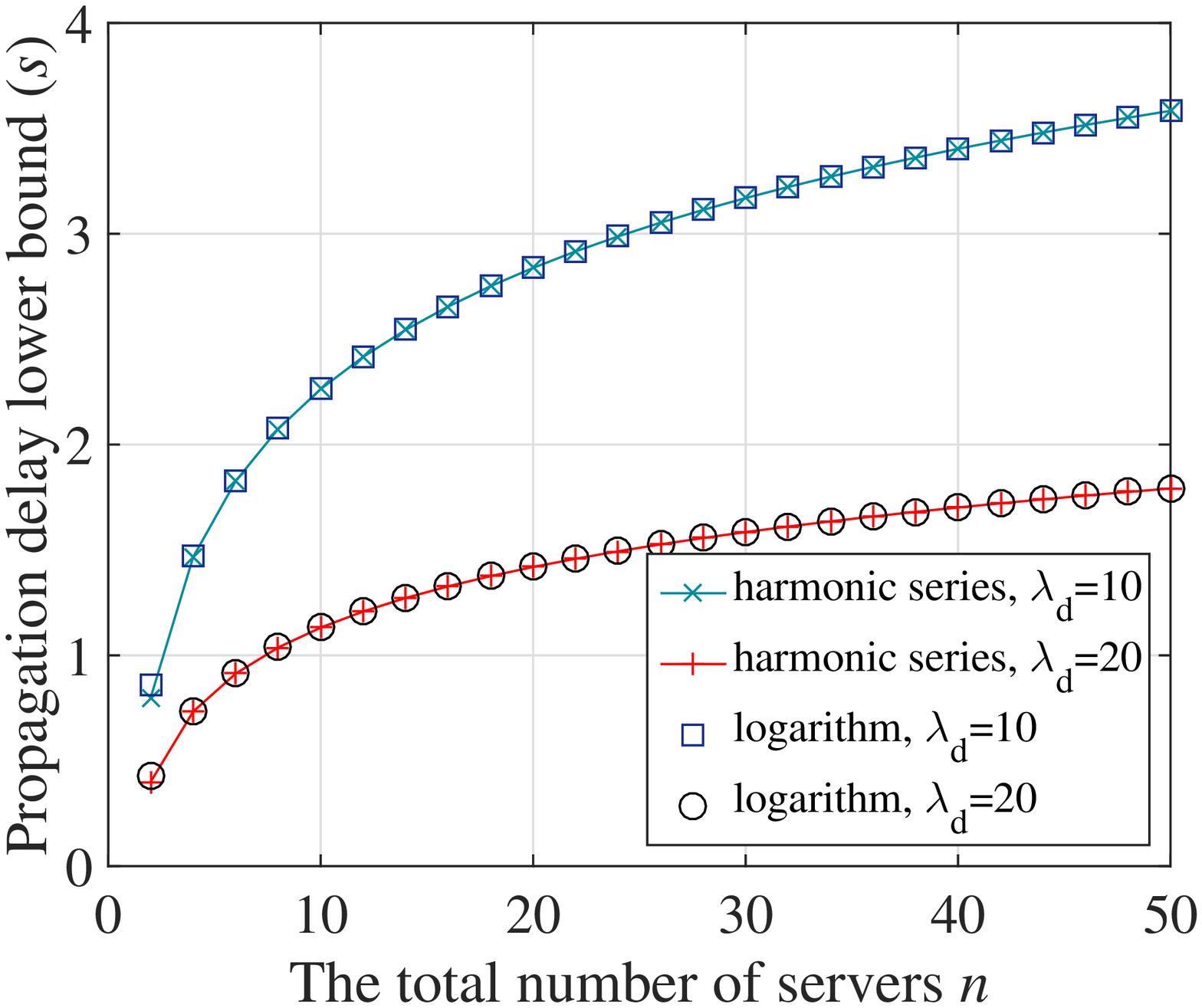}
}
\caption{{Block propagation performance evaluations.}}
\label{propagation}
\end{figure*}

In the first experiment, with fixed block generation time $t_b=10$ min and block size $s_b=1$ MB, we evaluate the impacts of the total number of servers $n$, the number of selected servers $k$, and the data rate in backhaul links $\lambda_d$ on block propagation performance.

\begin{table}[t]\small
\renewcommand\arraystretch{1.2}
\begin{center}
\caption{Parameter Settings}
\begin{tabular}{l|l}
\hline
\multicolumn{1}{l}{Parameter} & Value                                    \\
\hline
The total number of servers $n$             & $[10,50]$\\
The number of selected servers $k$          & $[1,n-1]$\\
The data rate in backhaul links $\lambda_d$ & $10$ Mbps, $20$ Mbps\\
Block generation time $t_b$		            & $[10^{-3},10^{1}]$ min \\
The size of a block $s_b$                   & $[10^{0},10^{4}]$ MB\\
The size of a transaction $s_t$             & $250$ bytes \\
The size of block header $s_h$		        & $80$ bytes\\
Transaction waiting time $t_w$	            & $10$ min\\
\hline
\end{tabular}
\end{center}
\end{table}

\emph{The number of informed servers vs. time} in Fig. \ref{propagation} (a) and (b) are obtained by equations (\ref{ETr}) and (\ref{cumulative}). It is shown that the growth curve of informed servers follows a double exponential behavior: an initial exponential growth phase in which the most of the servers will request the new block introduced by an \emph{inv} message, and an exponential convergence phase in which the most of the servers will ignore the \emph{inv} due to redundancy. This phenomenon indicates that the increasing rate of informed servers firstly increases and then decreases over time, which is consistent with the simulation results obtained by \cite{2-ChristianDecker}. With fixed $n=30$ and $\lambda_d=10$, Fig. \ref{propagation} (a) shows that $k$ can affect both the initial increasing rate and the convergence rate of informed servers, since a large $k$ results in a higher probability to select uninformed servers and a lower probability to enter failure state. On the other hand, with a fixed $k=4$, Fig. \ref{propagation} (b) shows that the curves for $n=10,20,30$ have the same initial increasing rate of informed servers, while a smaller $n$ will have a faster convergence rate. When network data rate $\lambda_d$ changes from $10$ Mbps to $20$ Mbps, the block transmission time in the network decreases, so we can see that the new block will be propagated to $90\%$ of total servers at a faster rate. After that, the increasing rate of the curve for $N=30,\lambda_d=20$ gets close to that for $N=30,\lambda_d=10$, due to a same failure probability shown in Fig. \ref{propagation} (f).

\begin{figure}[t]
\captionsetup{font={footnotesize}}
\centering
\subfigure[Forking probability vs. $t_b$]{
\includegraphics[width=4.1cm]{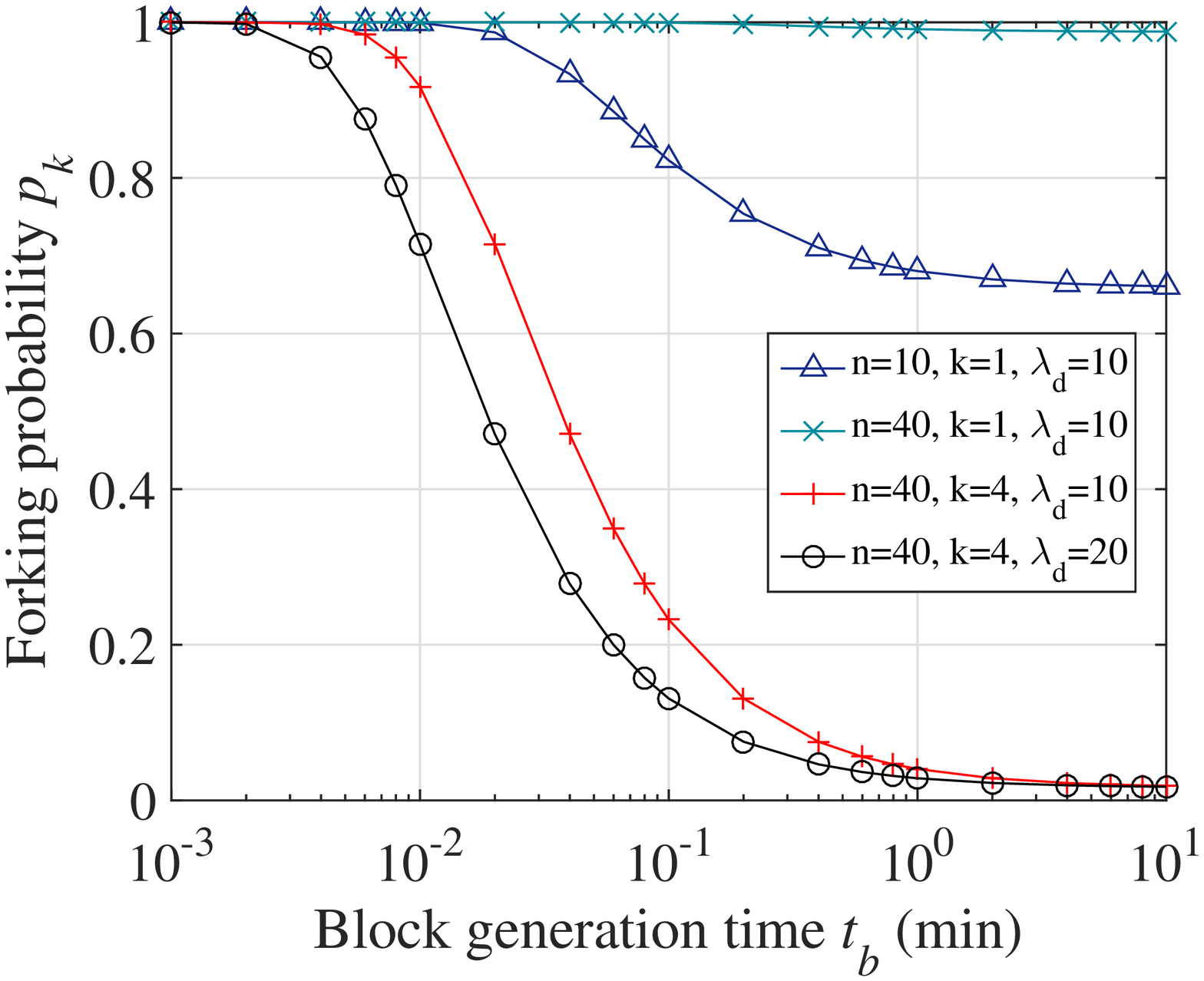}
}
\subfigure[Forking probability vs. $s_b$]{
\includegraphics[width=4.1cm]{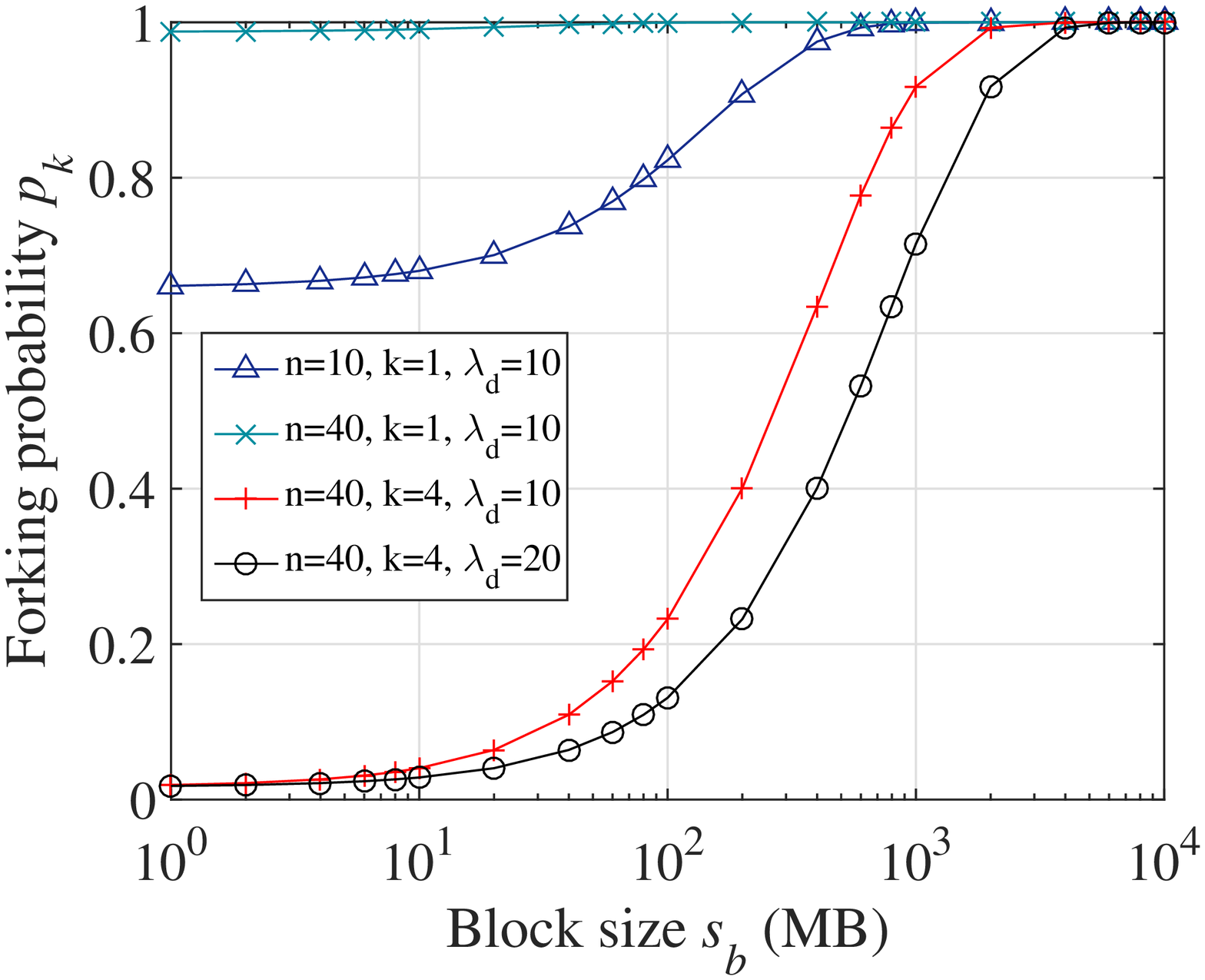}
}\\
\subfigure[Transaction throughput vs. $t_b$]{
\includegraphics[width=4.1cm]{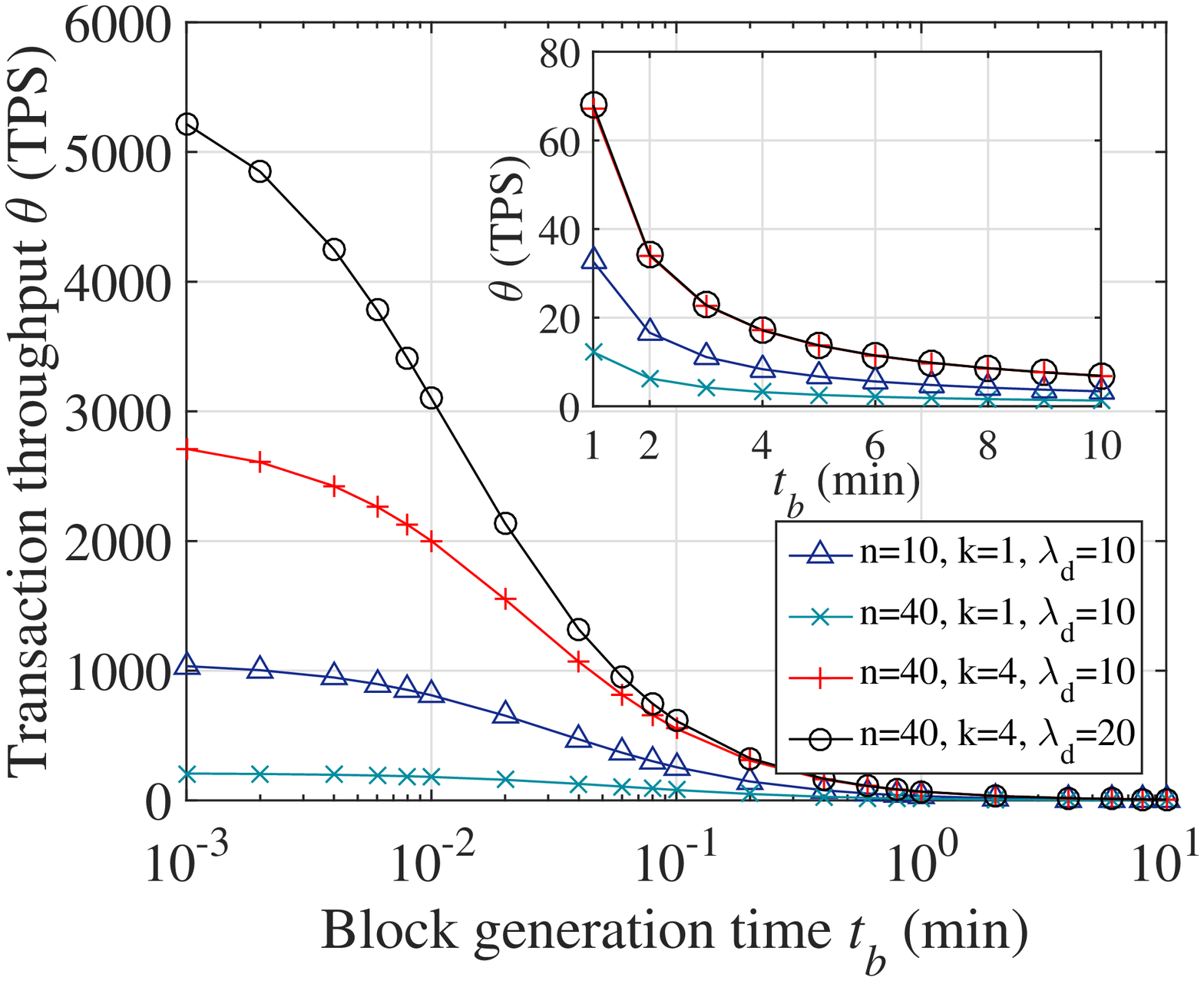}
}
\subfigure[Transaction throughput vs. $s_b$]{
\includegraphics[width=4.1cm]{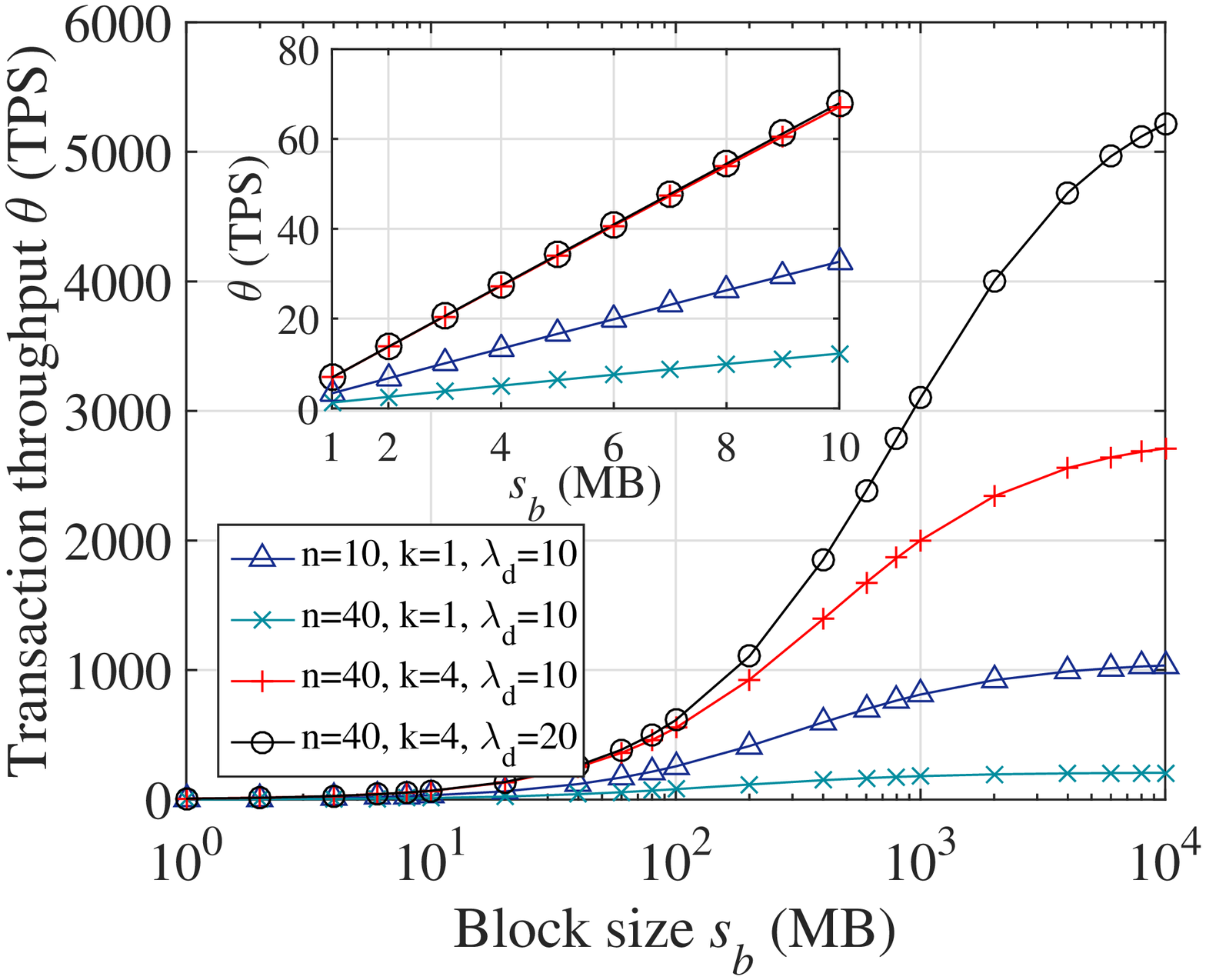}
}\\
\subfigure[Confirmation delay vs. $t_b$]{
\includegraphics[width=4.1cm]{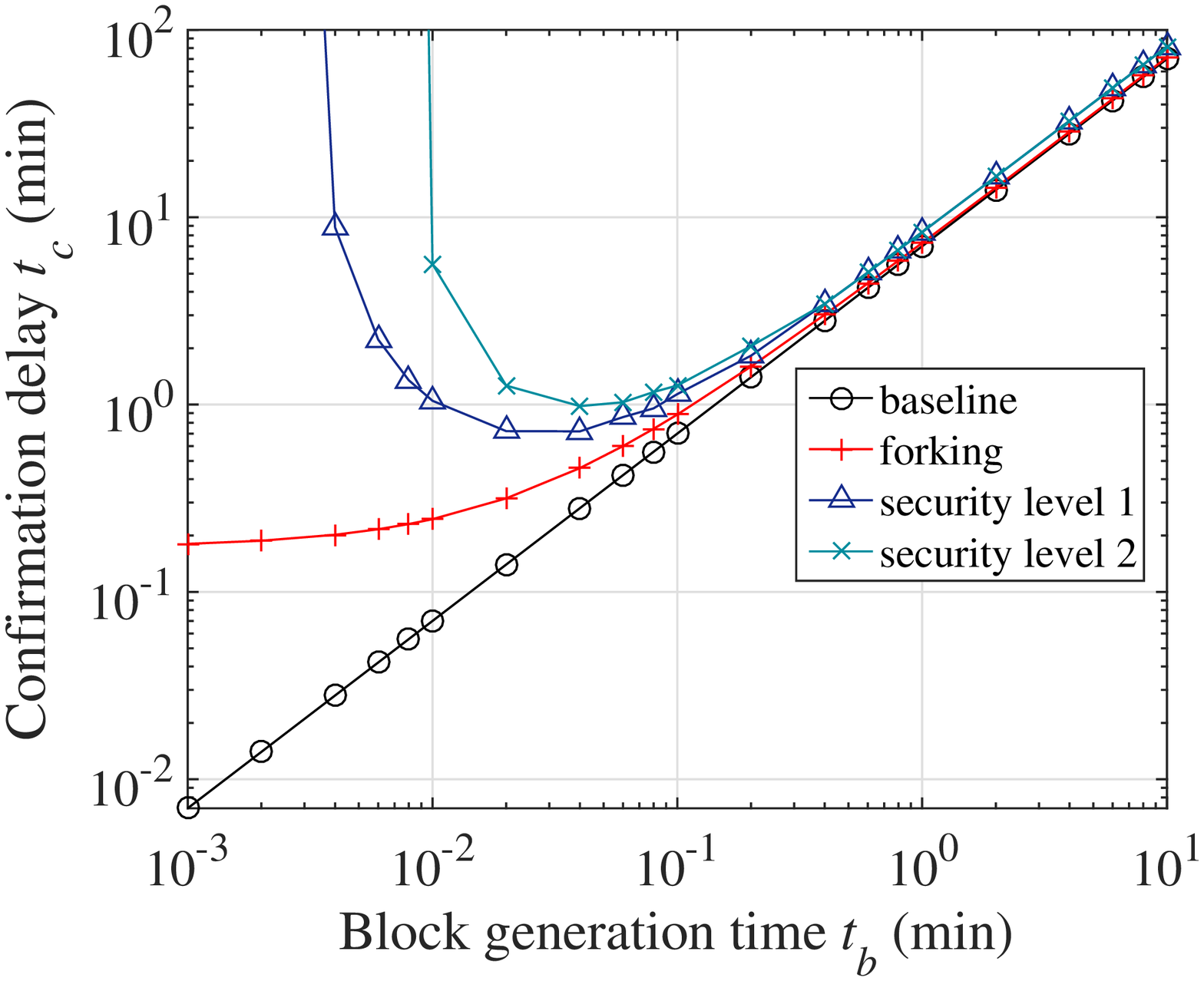}
}
\subfigure[Confirmation delay vs. $s_b$]{
\includegraphics[width=4.1cm]{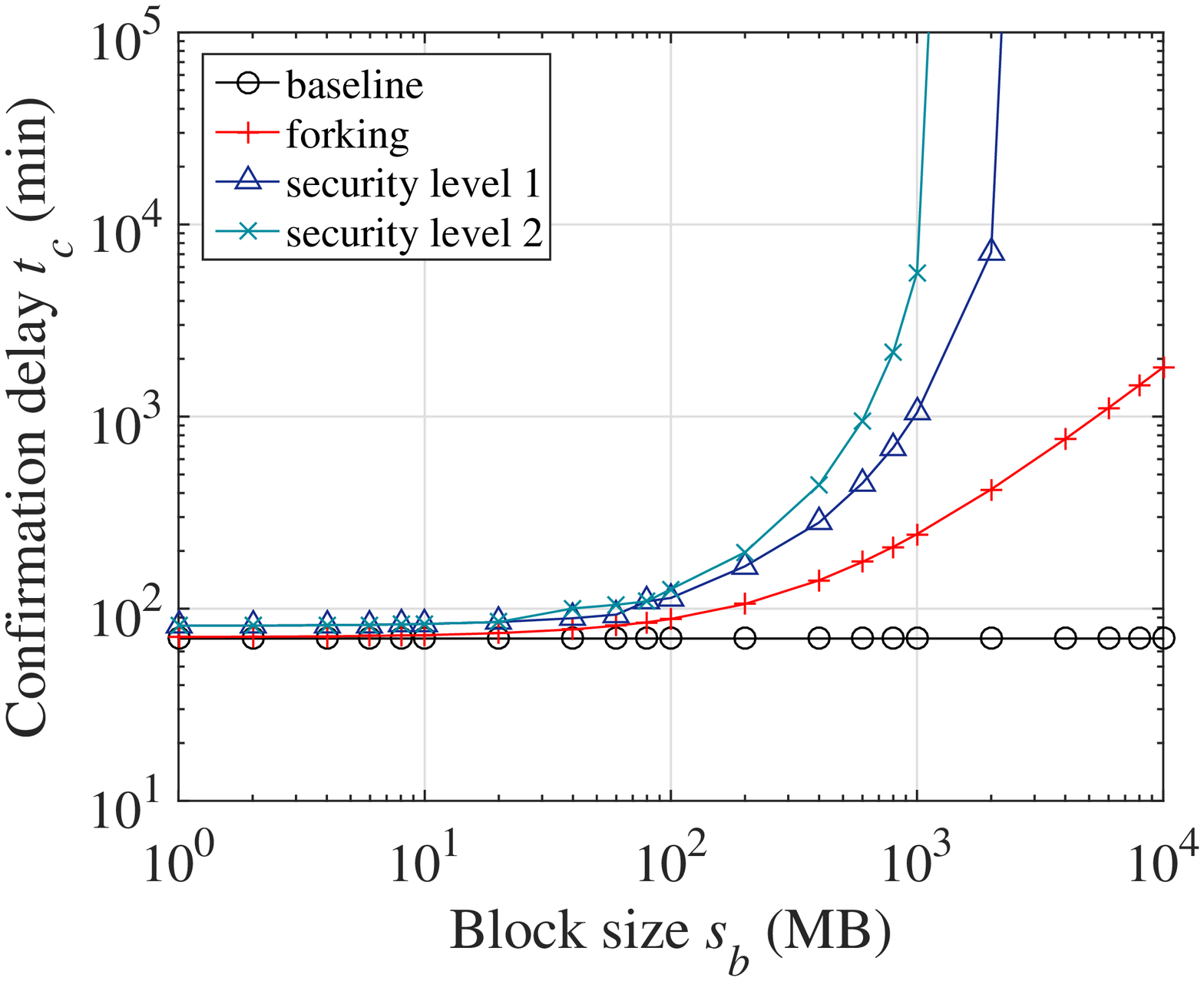}
}
\caption{{Forking probability and transaction-processing capability.}}
\label{efficiency}
\end{figure}

\begin{figure}[t]
\captionsetup{font={footnotesize}}
\centering
\subfigure[Fault tolerance vs. $t_b$]{
\includegraphics[width=4.1cm]{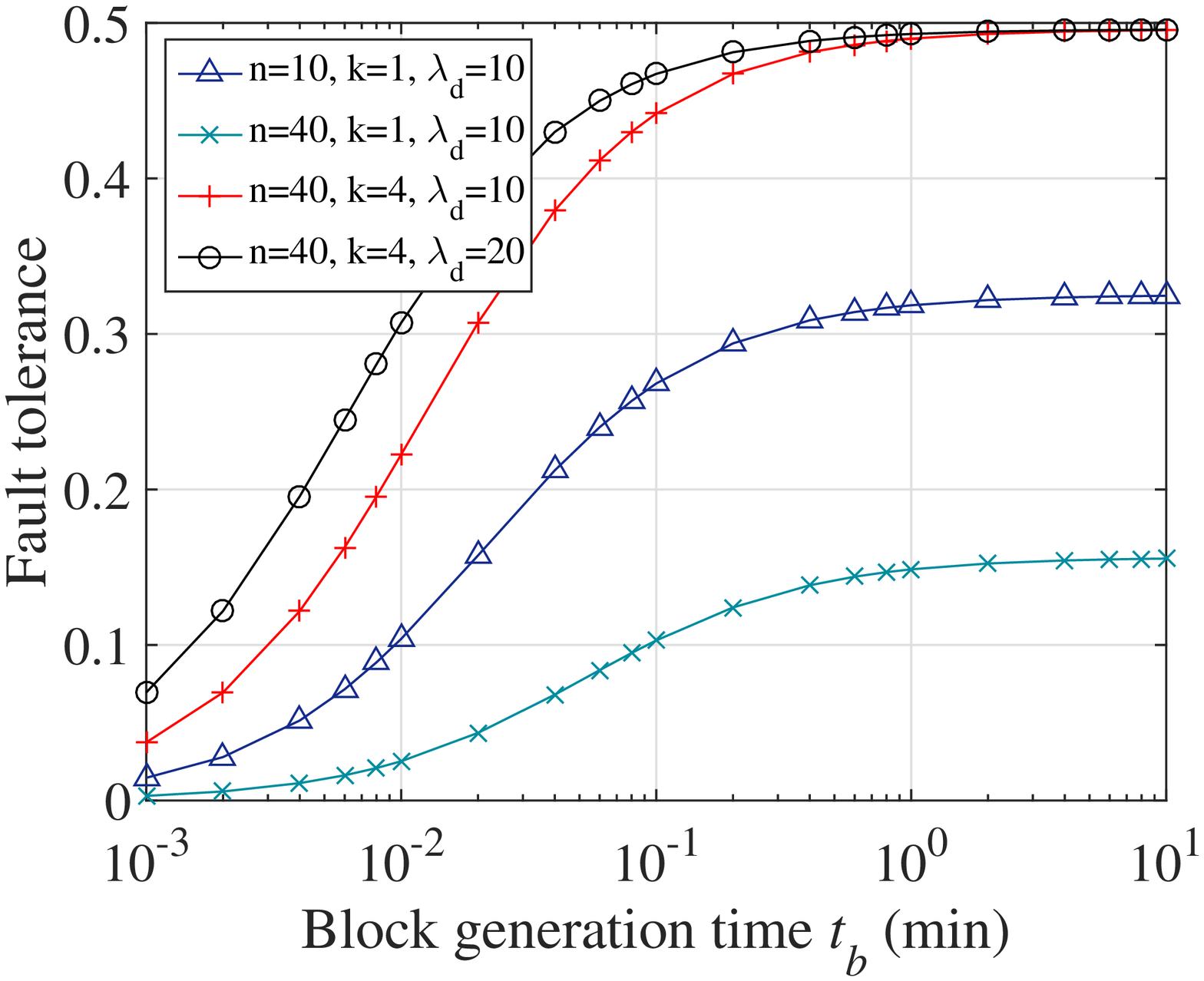}
}
\subfigure[Fault tolerance vs. $s_b$]{
\includegraphics[width=4.1cm]{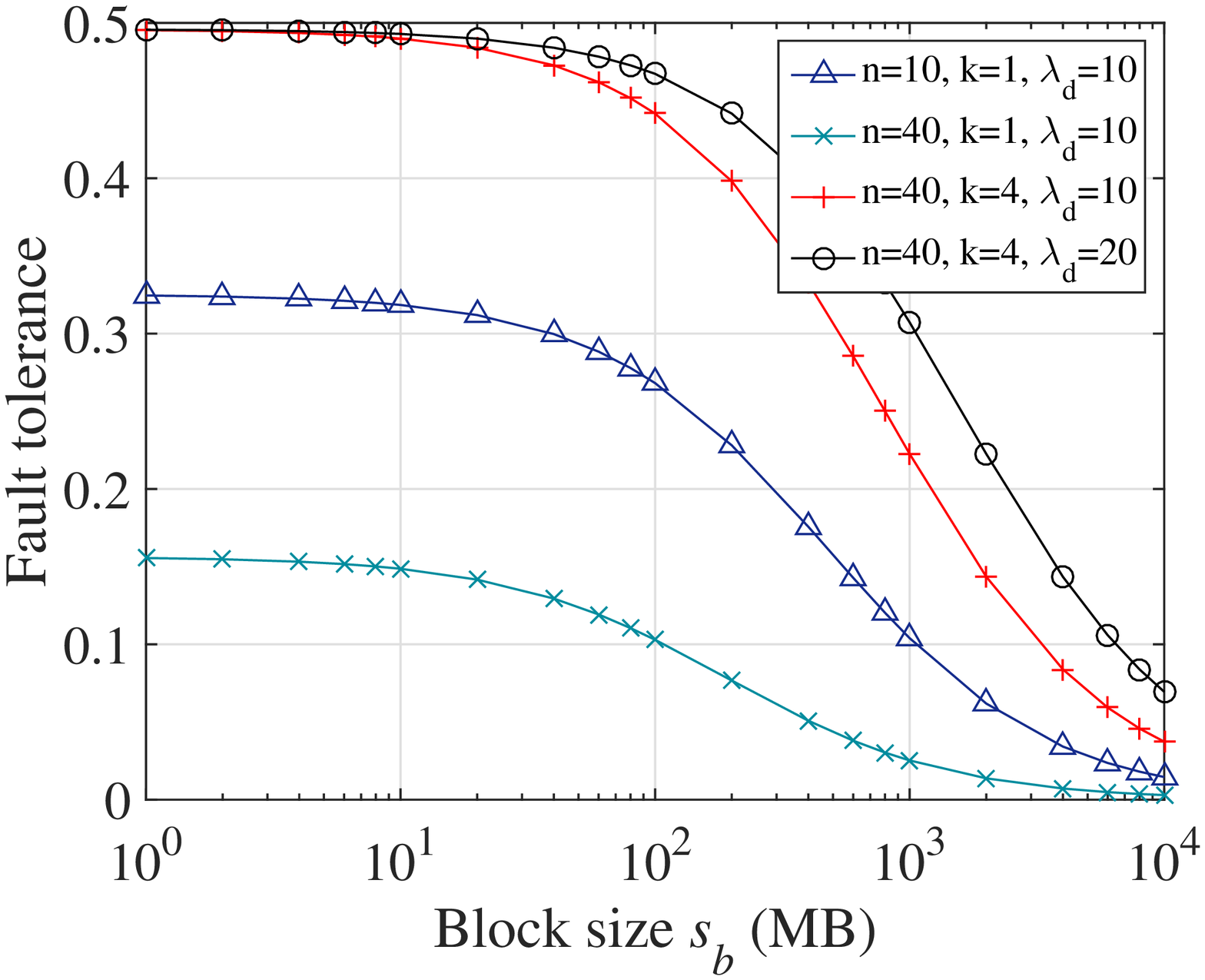}
}\\
\subfigure[Modification probability vs. $t_b$]{
\includegraphics[width=4.1cm]{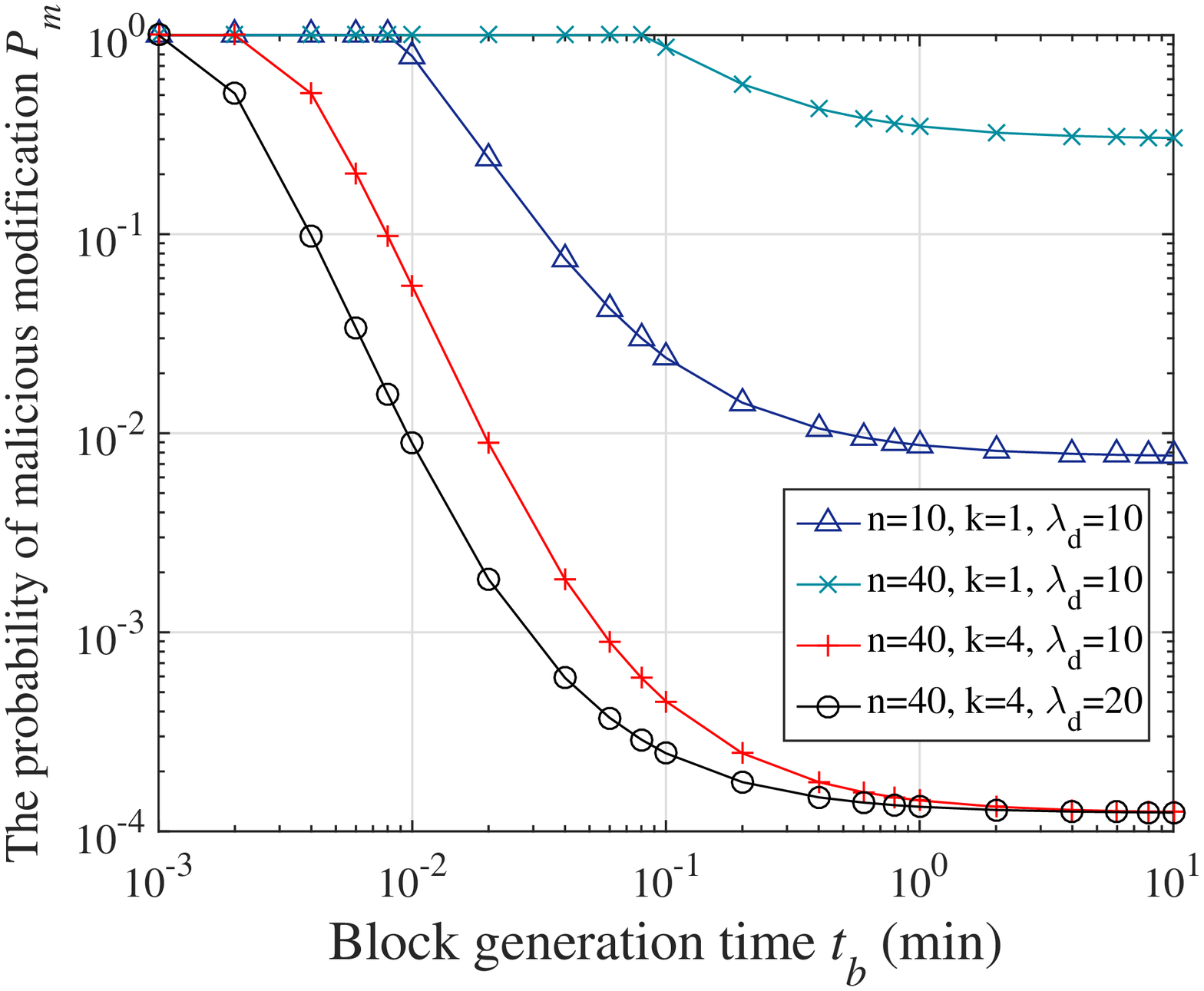}
}
\subfigure[Modification probability vs. $s_b$]{
\includegraphics[width=4.1cm]{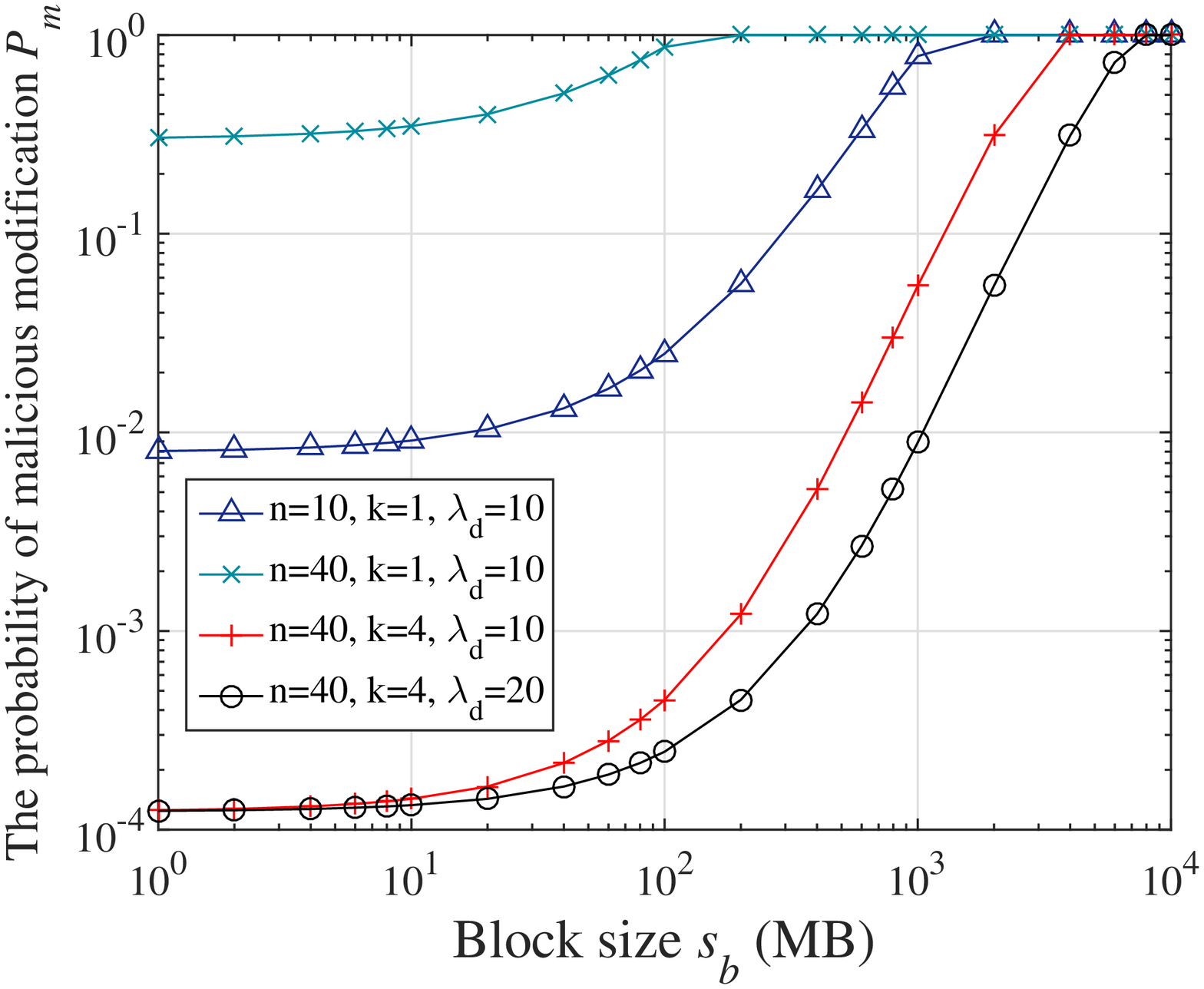}
}\\
\subfigure[Modification probability vs. $\lambda_m$]{
\includegraphics[width=4.1cm]{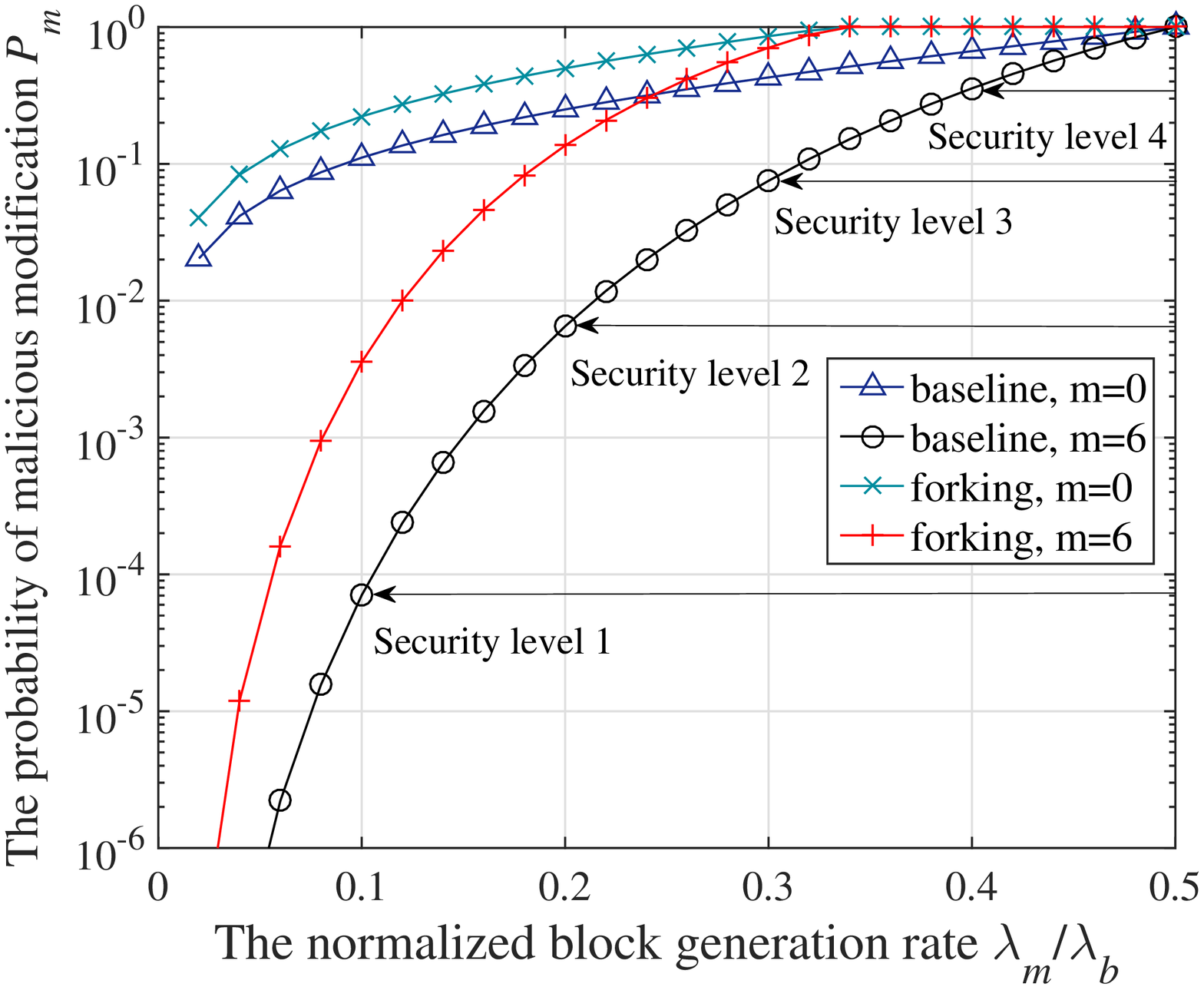}
}
\subfigure[Modification probability vs. $m$]{
\includegraphics[width=4.1cm]{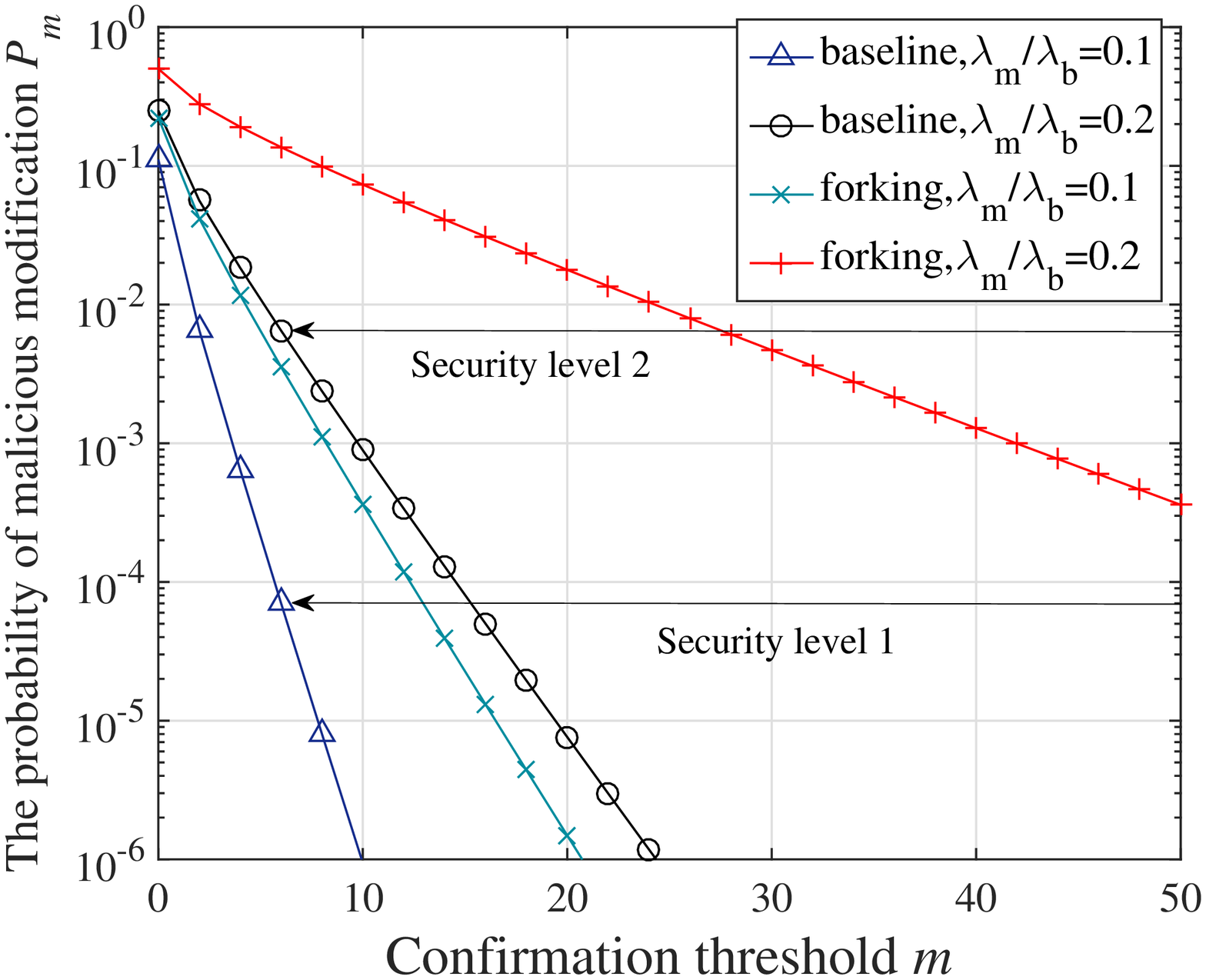}
}
\caption{{Blockchain security evaluations.}}
\label{security}
\end{figure}

\emph{Block propagation delay} in Fig. \ref{propagation} (c) and (d) are obtained by equation (\ref{Tp}). It is observed that block propagation delay increases monotonically with the total number of servers $n$, since a lager $n$ can result in more rounds for block transmissions and a higher failure probability. Compared with $n$, block propagation delay decreases monotonically with $k$ until reaching a lower bound. In fact, block propagation delay mainly consists of block transmission time in the network and block generation time due to propagation failure. When $k$ is small, block generation time in failure states is the dominant factor in block propagation delay, and each failure incurs an additional delay $t_b=600\text{s}$. When $k$ is large, the failure probability in Fig. \ref{propagation} (f) will be lower than $10^{-3}$, so that block transmission time becomes the dominant factor in block propagation delay. In this case, the block propagation delay in Fig. \ref{propagation} (d) stabilizes in the range of $[1,5]$s, which means that $k$ will have a lower impact on propagation delay as its value increases. Based on this phenomenon, we can minimize $k$ to reduce the communication cost of block propagation while approaching the delay lower bound with a given accuracy $\delta$. Fig. \ref{propagation} (g) gives the desired value of $k$ when $n\in[2,50]$, which is obtained by iterating $k$ from $1$ to $n-1$ to calculate equation (\ref{Tp}) until satisfying $t_p(k)-t_p(n-1)\leq \delta$.

The lower bound of propagation delay in Fig. \ref{propagation} (h) is obtained by equation (\ref{Tpbound}). It is clear to see that the results using harmonic series expression {\small$\frac{s_b}{\lambda_d}\sum\limits_{E_r=1}^{n-1}\frac{1}{E_r}$} match well with that using logarithmic expression {\small$\frac{s_b}{\lambda_d}\left[\ln(n-1)+\frac{1}
{2(n-1)}+\gamma\right]$}. The logarithmic expression demonstrates that the lower bound of propagation delay is determined by block size $s_b$, network data rate $\lambda_d$, and the total number of servers $n$. Optimizing these parameters will be an essential step in practical deployment of blockchain with delay requirement.

\emph{Propagation failure probability} in Fig. \ref{propagation} (e) and (f) are obtained by equation (\ref{pf}). It is shown that failure probability has an opposite trend with $n$ and $k$. The reason is that a larger $n$ can decrease the probability to select uninformed servers, namely $1-{C_{n-U_r-2}^{k} C_{U_r}^{0}}/{C_{n-2}^{k}}$, while larger $k$ can increase this probability. In other words, failure probability is determined by the difference between $n$ and $k$. The accuracy $\delta$ in Fig. \ref{propagation} (g) controls the difference between $n$ and $k$ to reduce failure probability, and thus propagation delay can approach its lower bound.

\subsection{Capability and Security Evaluations}

In the second experiment, we evaluate the impact of block generation time $t_b$ and block size $s_b$ on forking probability and capability-security metrics. Note that $t_b=10$ min and $s_b=1$ MB is the basic setting in Bitcoin, and we decrease $t_b$ from $10^{1}$ to $10^{-3}$ or increase $s_b$ from $10^{0}$ to $10^{4}$ for showing the upper/lower bounds and trade-offs of metrics.

\emph{Forking probability} in Fig. \ref{efficiency} (a) and (b) are obtained by equation (\ref{pk}). It is shown that forking probability will gradually rise to $1$, when we decrease $t_b$ or increase $s_b$. The symmetrical feature between Fig. \ref{efficiency} (a) and Fig. \ref{efficiency} (b) means that $t_b$ and $s_b$ has an opposite impact on forking probability. The reason is that forking probability is determined by the difference between block propagation delay and block generation time. Decreasing $t_b$ shortens the interarrival times of new blocks, while increasing $s_b$ prolongs block propagation delay, which result in a higher forking probability.

\emph{Transaction throughput} in Fig. \ref{efficiency} (c) and (d) are obtained by equation (\ref{theta}). The results show that transaction throughput can be improved by decreasing block generation time $t_b$ or increasing block size $s_b$. With the aggravation of forking, the increasing rate of transaction throughput will decline gradually, and the throughput will reach its upper bound when forking probability approaches $1$. It is shown that the upper bound of throughput is affected by the total number of servers $n$, the number of selected servers $k$, network data rate $\lambda_d$, which is consistent with the analysis for equation (\ref{thetaupper}). Based on the results, $t_b$ and $s_b$ can be well configured to improve throughput when forking probability is low. On the other hand, when forking probability approaches $1$, $t_b$ and $s_b$ will have a low impact on throughput. In this case, it is essential to optimize network data rate $\lambda_d$ or reduce the difference between $n$ and $k$.

\emph{Fault tolerance} in Fig. \ref{security} (a) and (b) are obtained by equation (\ref{tolerance}). Based on (\ref{tolerance}), fault tolerance is inversely proportional to the number of forks in the network, and thus a high forking probability can result in a low fault tolerance. It is shown that fault tolerance will drop to a value below $0.1$ when $t_b=10^{-3}$ or $s_b=10^{4}$, which means that the malicious server can modify arbitrary data items with $10\%$ of total consensus resources. On the other hand, we know that throughput can be improved to thousands of TPS when $t_b=10^{-3}$ or $s_b=10^{4}$. The relationship between transaction throughput and fault tolerance demonstrates a trade-off of blockchain: a high transaction throughput can be achieved by adjusting block generation time or block size, but it sacrifices the fault tolerance of blockchain.

\emph{The probability of malicious modification} in Fig. \ref{security} (c)-(f) are obtained by equations (\ref{pm}) and (\ref{pw}). Fig. \ref{security} (c) and (d) show that the results of probability $p_m$ have a similar trend with that of forking probability, which indicates that $p_m$ might be affected forking probability. To verify this observation, Fig. \ref{security} (e) compares the probability $p_m$ under the impact of forking with that of baseline (an ideal condition without forking). It is shown that forking can not only increase the probability of malicious modification, but also affect the convergence point of $p_m=1$. In other words, malicious servers can guarantee the success of attack using much less consensus resources when forking probability is high.

One way to reduce the adverse impact of forking on security is to adjust confirmation threshold $m$. Fig. \ref{security} (e) reports the probability $p_m$ when confirmation threshold $m$ varies from $0$ to $50$. Note that $m=0$ represents that a data item is not encrypted by data producer, so the malicious server can obtain the data without confirmation; $m=6$ is the basic setting in Bitcoin, namely that a data item having $6$ confirmation is considered to be irreversible. Without forking, the probability $p_m$ for $m=6$ can be defined as the security level in Bitcoin, where security level $1$ refers to the probability $p_m$ when the malicious server has $10\%$ of total block generation rate; security level $2$ refers to the probability $p_m$ when the malicious server has $20\%$ of total block generation rate. Under the impact of forking, the probability $p_m$ reaches security level $1$ when $m=16$, and it reaches security level $2$ when $m=28$. It means that the adverse impact of forking on security can be reduced by waiting for more block confirmations.

\emph{Confirmation delay} in Fig. \ref{efficiency} (e) and (f) are obtained by equation (\ref{tc}). Without forking, confirmation delay decreases monotonically as $t_b$ decreases, and it is independent of $s_b$. Under forking, confirmation delay will reach its lower bound as $t_b$ decreases, which matches with the analysis for equation (\ref{tclower}). Compared with $t_b$, confirmation delay will increase exponentially with $s_b$ because of forking. Moreover, we can see that security level $1$ and $2$ can incur additional delay, due to the change of confirmation threshold $m$. This result demonstrates that a high security level is achieved by sacrificing the delay performance of blockchain.

\subsection{Properties of Failure State}

In the last experiment, we evaluate the transition probability of failure state and the expected time to leave a failure state based on equations (\ref{second}), (\ref{event}), and (\ref{T0}). Fig. \ref{massfunction} (a) compares the transition probabilities of failure states $\{10,0\}$, $\{20,0\}$, and $\{30,0\}$. It is shown that the above failure states will transit to states $\{16,0\}$, $\{24,0\}$, and $\{32,0\}$ with the highest probability after the next block propagation. A smaller failure state has a faster transition rate since there are more uninformed servers. Compare Fig. \ref{massfunction} (a) with (b), we can see that the transition rate of failure state increases when $k$ changes from $1$ to $2$, because $k$ increases the probability to select uninformed servers.

Fig. \ref{massfunction} (c) and (d) show the probability that a new block is generated in failure state. Compared with the results for $k=2$, the new block has a higher probability to be generated before the failure state $\{I_r,0\}$ when $k=1$. This means that $\{I_r,0\}$ can transit to the next state without waiting for the generation of new block, and thus the Markov chain can leave the failure state in a short time. On the other hand, $\{I_r,0\}$ should wait for the new block when $k=2$ due to the probability approaches $1$, which spends more time to leave the failure state. Therefore, we can see that the time to leave a failure state mainly depends on $k$ in Fig. \ref{massfunction} (e) and (f). When $k$ is small, the time first decreases and then increases with $I_r$; when $k$ is sufficiently large (say larger than $4$ for $n\leq 40$), the time to leave a failure state approaches block generation time $t_b$.

\section{Conclusions}

In this paper, we have studied the performance trade-offs and theoretical bounds of blockchain from the perspective of block propagation and forking problem. The block propagation process has been modeled as Markov chain to capture the impact of asynchronous block transmissions in the network. Based on Markov chain, we have derived the closed-form expressions of blockchain performance metrics, including block propagation performance and blockchain capability-security performance. The numerical results quantify the trade-off between transaction throughput and fault tolerance, as well as the trade-off between confirmation delay and modification probability. It is shown that transaction throughput can be improved from seven TPS to thousands of TPS in a consensus domain, and its upper bound is highly affected by network parameters, i.e., the total number of servers, the number of selected servers, and network data rate. At the point when throughput equals to one thousand TPS, the sacrificed fault tolerance ranges from $10\%$ to $40\%$, depending on forking probability. Meanwhile, adjusting confirmation threshold will logarithmically reduce modification probability (a larger threshold has a lower impact on modification probability), at the cost of increasing confirmation delay.


\begin{thebibliography}{11}


\bibitem{5-RuizheYang}
R. Yang, F. R. Yu, P. Si, Z. Yang, and Y. Zhang,
\newblock {``Integrated blockchain and edge computing systems: a survey, some research issues and challenges,''}
\newblock {\em IEEE Commun. Surveys Tuts.}, vol. 21, no. 2, pp. 1508-1532, 2nd Quart. 2019.

\bibitem{6-MengShen}
M. Shen, J. Duan, L. Zhu, J. Zhang, X. Du, and M. Guizani,
\newblock {``Blockchain-based incentives for secure and collaborative data sharing in multiple clouds,''}
\newblock {\em IEEE J. Sel. Areas Commun.}, vol. 38, no. 6, pp. 1229-1241, Jun. 2020.


\bibitem{6-YueqiangXu}
Y. Xu, H. Zhang, H. Ji, L. Yang, X. Li, and V. C. M. Leung,
\newblock {``Transaction throughput optimization for integrated blockchain and MEC system in IoT,''}
\newblock {\em IEEE Trans. Wireless Commun.}, vol. 21, no. 2, pp. 1022-1036, Feb. 2022.

\bibitem{4-YaodongHuang}
Y. Huang, Y. Zeng, F. Ye, and Y. Yang,
\newblock {``Fair and protected profit sharing for data trading in pervasive edge computing environments,''}
\newblock In {\em Proc. IEEE Int. Conf. Comput. Commun. (INFOCOM)}, Toronto, Canada, Jun. 2020, pp. 1718-1727.

\begin{figure}[t]
\captionsetup{font={footnotesize}}
\centering
\subfigure[State after the next propagation]{
\includegraphics[width=4.1cm]{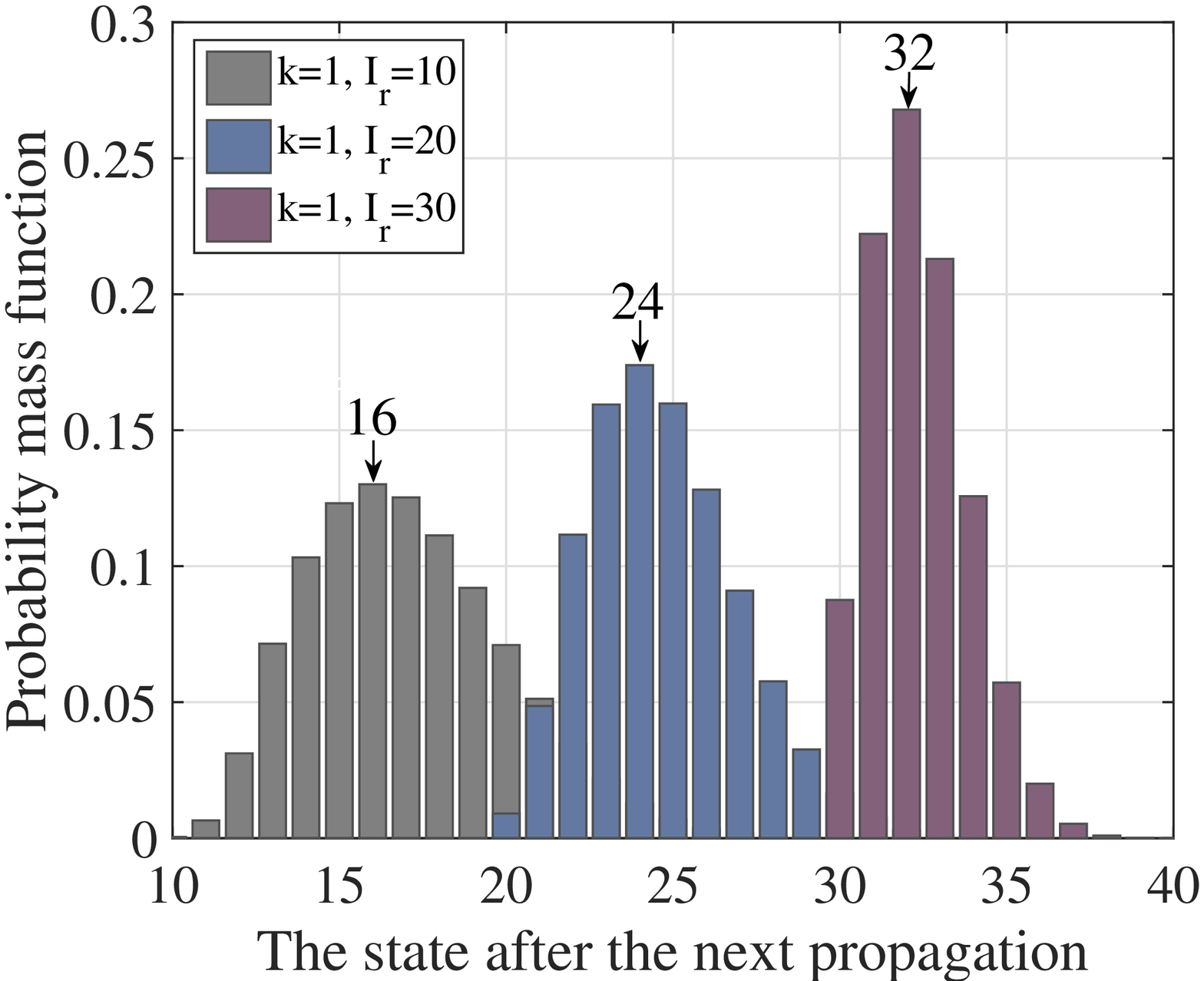}
}
\subfigure[State after the next propagation]{
\includegraphics[width=4.1cm]{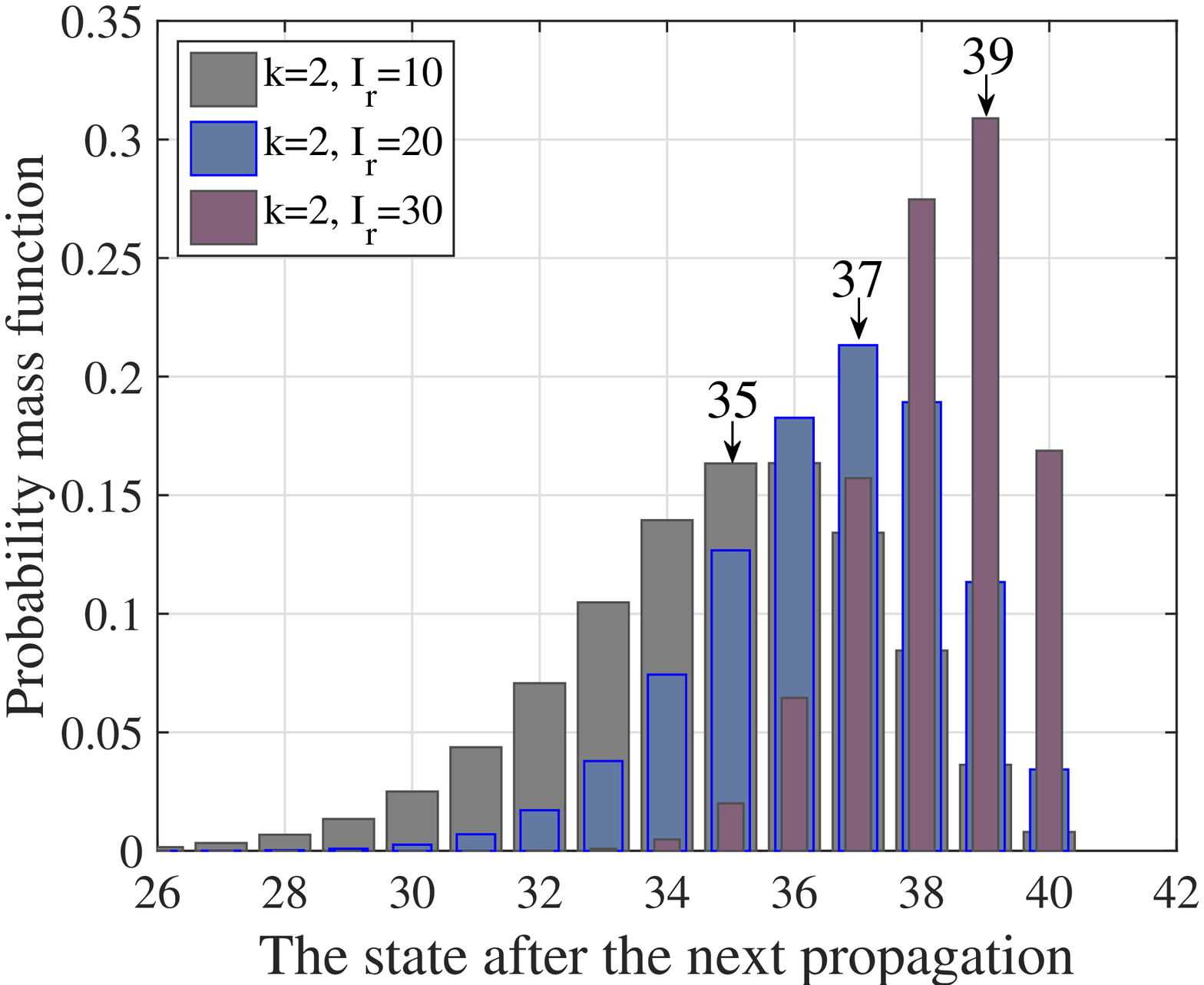}
}\\
\subfigure[The state to generate new block]{
\includegraphics[width=4.1cm]{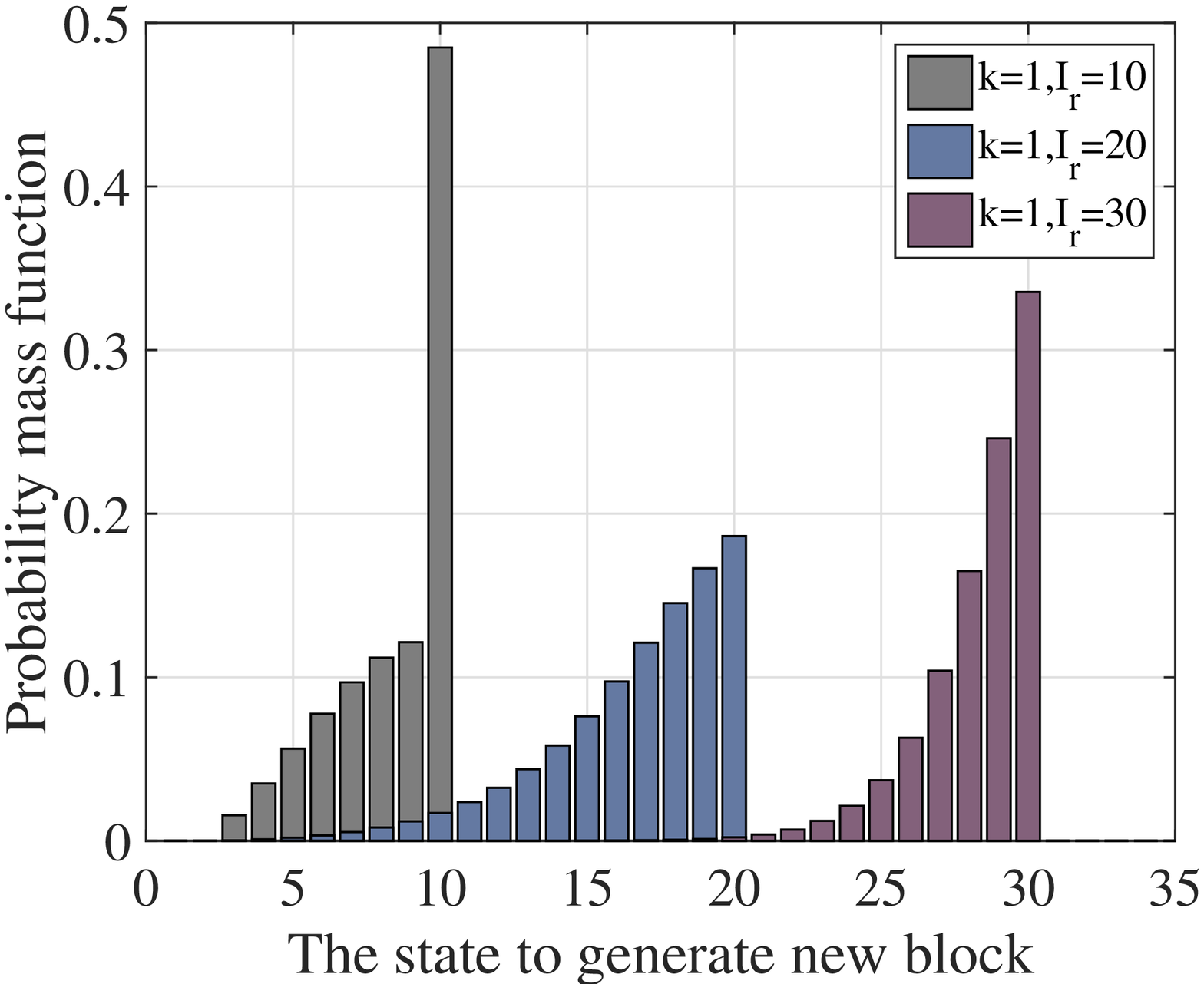}
}
\subfigure[The state to generate new block]{
\includegraphics[width=4.1cm]{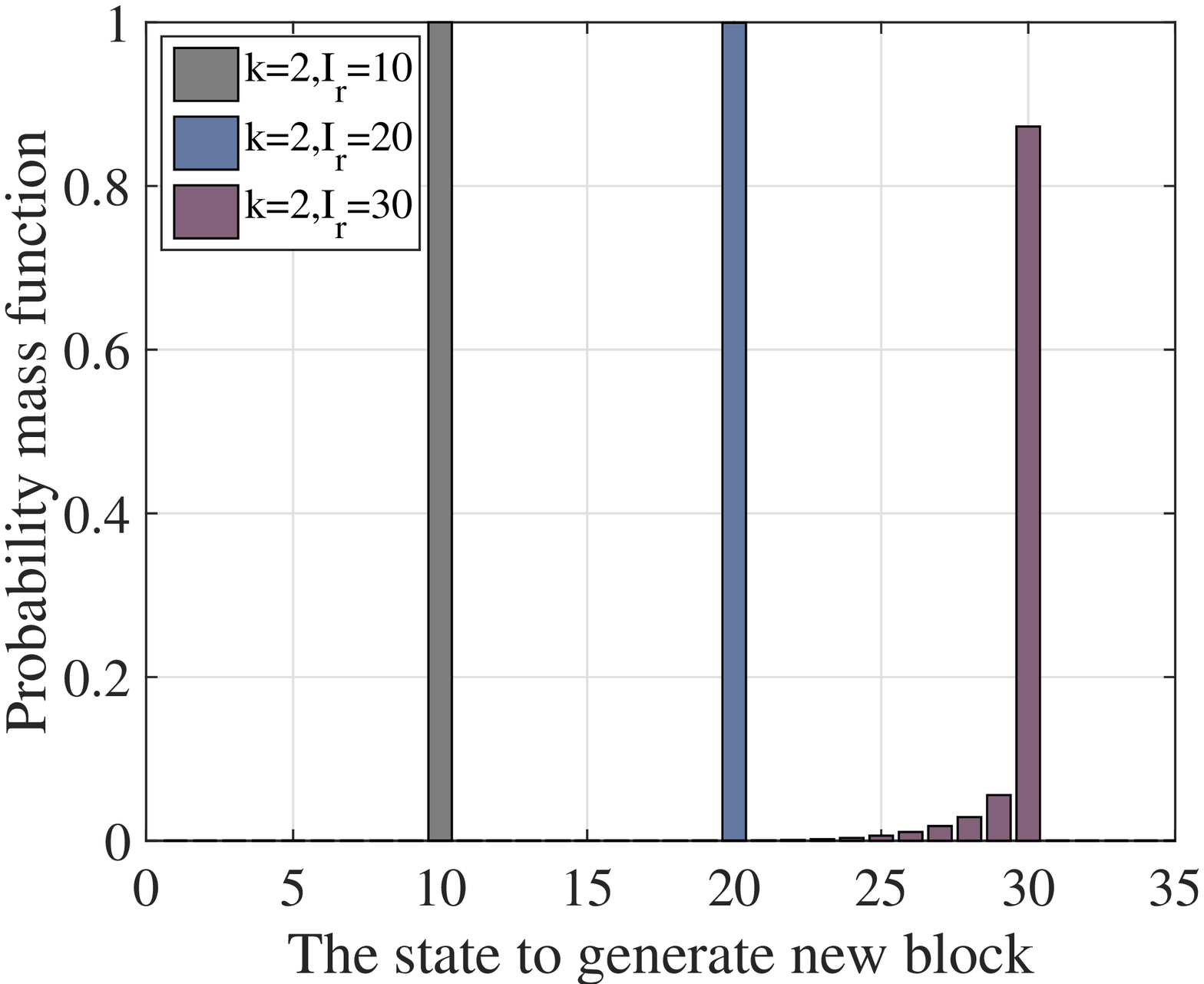}
}\\
\subfigure[The time to leave a failure state]{
\includegraphics[width=4.1cm]{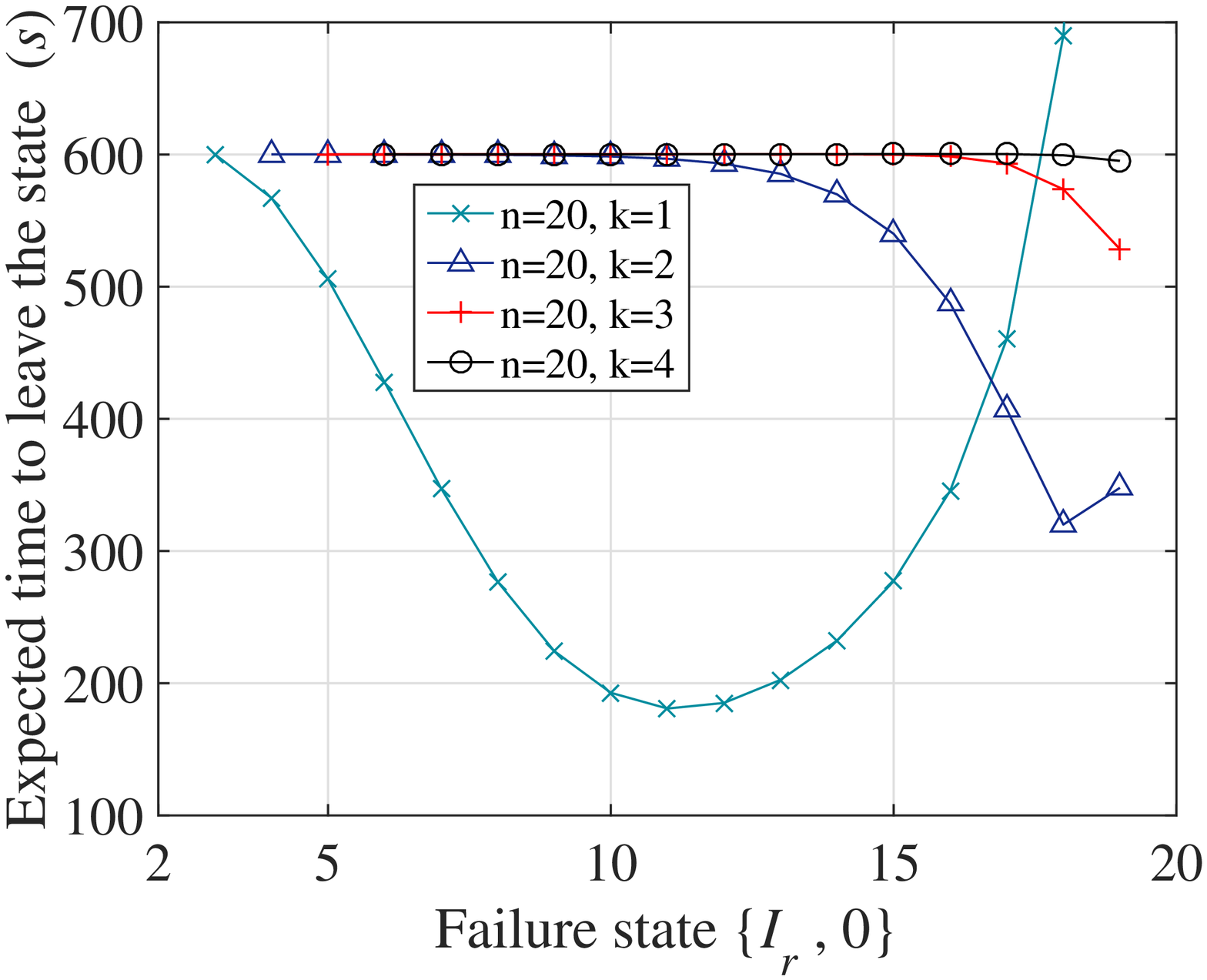}
}
\subfigure[The time to leave a failure state]{
\includegraphics[width=4.1cm]{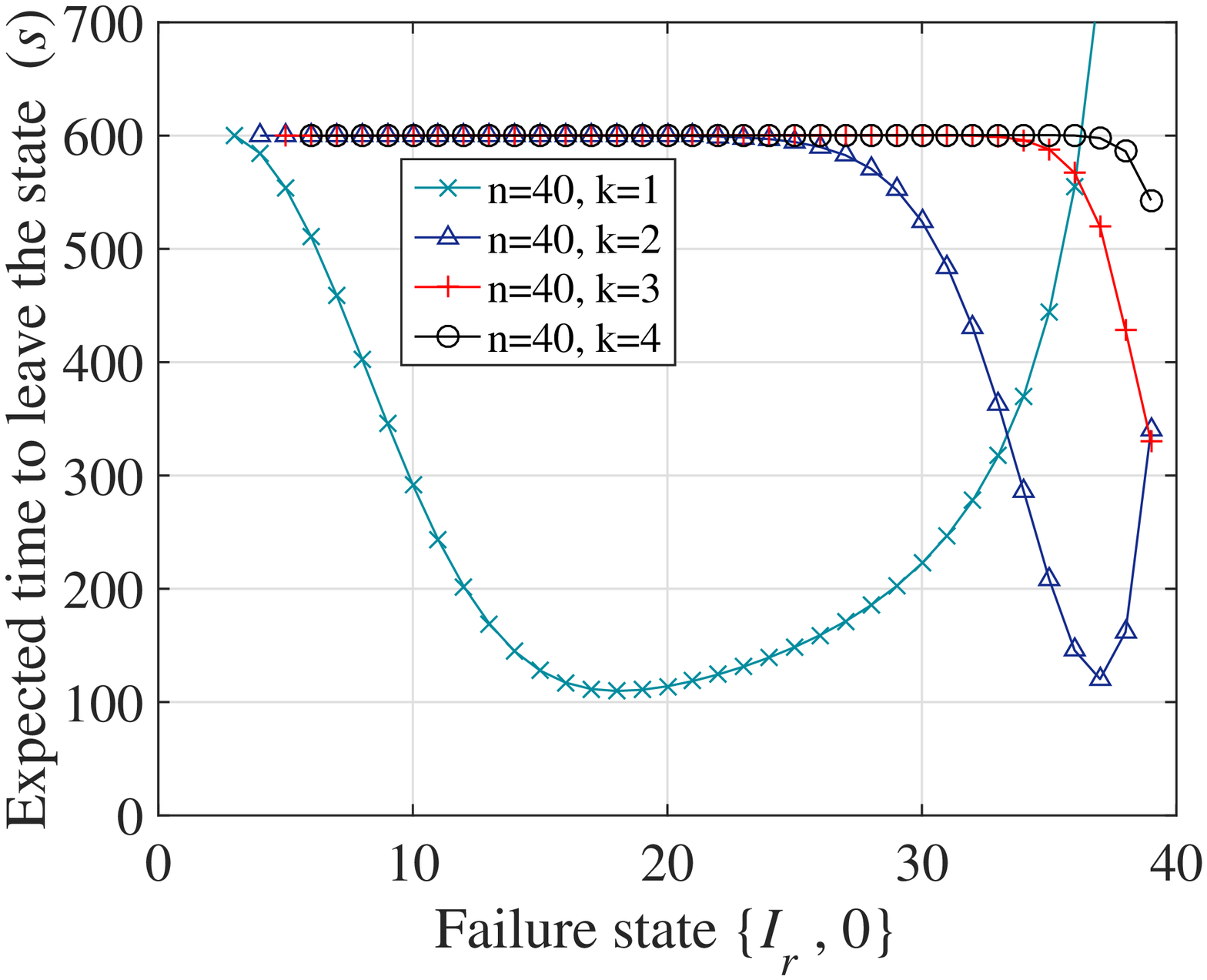}
}
\caption{{Probability mass function and the time to leave a failure state}}
\label{massfunction}
\end{figure}

\bibitem{6-Jiawen}
J. Kang, Z. Xiong, D. Niyato, D. Ye, D. I. Kim, and J. Zhao,
\newblock {``Toward secure blockchain-enabled internet of vehicles: optimizing consensus management using reputation and contract theory,''}
\newblock {\em IEEE Trans. Veh. Technol.}, vol. 68, no. 3, pp. 2906-2920, Mar. 2019.


\bibitem{1-3GPP}
3GPP,
\newblock {``Technical specification group services and system aspects: service requirements for the 5G system,''} Technical specification, Dec. 2021. [Online]. Available: https://www.3gpp.org.

\bibitem{7-JiawenKang}
J. Kang \textit{et al.},
\newblock {``Blockchain for secure and efficient data sharing in vehicular edge computing and networks,''}
\newblock {\em IEEE Internet of Things J.}, vol. 6, no. 3, pp. 4660-4670, Jun. 2019.

\bibitem{7-LiangXiao}
L. Xiao \textit{et al.},
\newblock {``A reinforcement learning and blockchain-based trust mechanism for edge networks,''}
\newblock {\em IEEE Trans. Commun.}, vol. 68, no. 9, pp. 5460-5470, Sept. 2020.

\bibitem{5-ZheYang}
Z. Yang, K. Yang, L. Lei, K. Zheng, V. C. M. Leung,
\newblock {``Blockchain-based decentralized trust management in vehicular networks,''}
\newblock {\em IEEE Internet of Things J.}, vol. 6, no. 2, pp. 1495-1505, Apr. 2019.

\bibitem{4-HaoyeChai}
H. Chai, S. Leng, Y. Chen, and K. Zhang,
\newblock {``A hierarchical blockchain-enabled federated learning algorithm for knowledge sharing in internet of vehicles,''}
\newblock {\em IEEE Trans. Intell. Transp. Syst.}, vol. 22, no. 7, pp. 3975-3986, Jul. 2021.

\bibitem{6-DekeGuo}
D. Guo, J. Xie, X. Shi, H. Cai, C. Qian, and H. Chen,
\newblock {``HDS: a fast hybrid data location service for hierarchical mobile edge computing,''}
\newblock {\em IEEE/ACM Trans. Netw.}, vol. 29, no. 3, pp. 1308-1320, Jun. 2021.

\bibitem{5-JianChang}
J. Chang, J. Ni, J. Xiao, X. Dai, and H. Jin,
\newblock {``Synergychain: a multichain-based data sharing framework with hierarchical access control,''}
\newblock {\em IEEE Internet of Things J.}, to be published.

\bibitem{5-PinarOzisik}
A. Ozisik, B. Levine, G. Bissias, G. Andresen, D. Tapp, and S. Katkuri,
\newblock {``Graphene: efficient interactive set reconciliation applied to blockchain propagation,''}
\newblock In {\em Proc. ACM Special Int. Group Data Commun. (SIGCOMM)}, Beijing, China, Aug. 2019, pp. 303-317.

\bibitem{2-ChristianDecker}
C. Decker, and R. Wattenhofer,
\newblock {``Information propagation in the Bitcoin network,''}
\newblock In {\em Proc. 13th IEEE Int. Conf. Peer-to-Peer Comput. (P2P)}, Trento, Italy, Sept. 2013, pp. 1-10.

\bibitem{5-Jelena}
J. Misic, V. B. Misic, X. Chang, S. G. Motlagh, and M. Z. Ali,
\newblock {``Modeling of Bitcoin's blockchain delivery network, ''}
\newblock {\em IEEE Trans. Netw. Sci. Eng.}, vol. 7, no. 3, pp. 1368-1381, Jul.-Sep. 2020.

\bibitem{3-YahyaShahsavari}
Y. Shahsavari, K. Zhang, and C. Talhi,
\newblock {``A theoretical model for block propagation analysis in Bitcoin network,''}
\newblock {\em IEEE Trans. Eng. Manag.}, to be published.

\bibitem{1-Nakamoto}
S. Nakamoto,
\newblock {``Bitcoin: a peer-to-peer electronic cash system,''}
White paper, 2009. [Online]. Available: https://bitcoin.org/bitcoin.pdf.

\bibitem{7-ChenhanXu}
C. Xu \textit{et al.},
\newblock {``Making big data open in edges: a resource-efficient blockchain-based approach,''}
\newblock {\em IEEE Trans. Parallel Distrib. Syst.}, vol. 30, no. 4, pp. 870-882, Apr. 2019.

\bibitem{6-JiananLi}
J. Li, J. Wu, J. Li, A. K. Bashir, M. J. Piran, and A. Anjum,
\newblock {``Blockchain-based trust edge knowledge inference of multi-robot systems for collaborative tasks,''}
\newblock {\em IEEE Commun. Mag.}, vol. 59, no. 7, pp. 94-100, Jul. 2021.

\bibitem{5-YueyueDai}
Y. Dai, D. Xu, K. Zhang, S. Maharjan, and Y. Zhang,
\newblock {``Deep reinforcement learning and permissioned blockchain for content caching in vehicular edge computing and networks,''}
\newblock {\em IEEE Trans. Veh. Technol.}, vol. 69, no. 4, pp. 4312-4324, Apr. 2020.

%

\bibitem{4-JinHuaChen}
J. Chen, M. Chen, G. Zeng, and J. Weng,
\newblock {``BDFL: a Byzantine-fault-tolerance decentralized federated learning method for autonomous vehicle,''}
\newblock {\em IEEE Trans. Veh. Technol.}, vol. 70, no. 9, pp. 8639-8652, Sept. 2021.

\bibitem{6-JinmingShi}
J. Shi, J. Du, Y. Shen, J. Wang, J. Yuan, and Z. Han,
``DRL-based V2V computation offloading for blockchain-enabled vehicular networks,''
\newblock {\em IEEE Trans. Mobile Comput.}, to be published.

\bibitem{7-LiangYuan}
L. Yuan \textit{et al.},
\newblock {``CSEdge: enabling collaborative edge storage for multi-access edge computing based on blockchain,''}
\newblock {\em IEEE Trans. Parallel Distrib. Syst.}, vol. 33, no. 8, pp. 1873-1887, Aug. 2022.


\bibitem{4-YangXiao}
Y. Xiao, N. Zhang, W. Lou, and Y. T. Hou,
\newblock {``A survey of distributed consensus protocols for blockchain networks,''}
\newblock {\em IEEE Commun. Surveys Tuts.}, vol. 22, no. 2, pp. 1432-1465, 2nd Quart. 2020.


\bibitem{1-Antonopoulos}
A. M. Antonopoulos,
\newblock {``Mastering Bitcoin: unlocking digital cryptocurrencies,''} 2nd ed. Sebastopol, CA, USA: O'Reilly Media, Inc., Jun. 2017.

\bibitem{3-Chen}
D. C. Chen, T. Q. S. Quek, and M. Kountouris,
\newblock {``Backhauling in heterogeneous cellular networks: modeling and tradeoffs,''}
\newblock {\em IEEE Trans. Wireless Commun.}, vol. 14, no. 6, pp. 3194-3206, Jun. 2015.

\bibitem{1-Ross}
S. M. Ross,
\newblock {``Introduction to probability models,''} 11th ed. New York, NY, USA: Academic, 2014.

\bibitem{1-Havil}
J. Havil,
\newblock {``The harmonic series,''} Ch. 2 in Gamma: exploring Euler's constant. Princeton, NJ: Princeton Univ. Press, pp. 21-25, 2003.

\bibitem{7-yixinli}
Y. Li \textit{et al.},
\newblock ``Direct acyclic graph-based ledger for Internet of Things: performance and security analysis,''
\newblock {\em IEEE/ACM Trans. Netw.}, vol. 28, no. 4, pp. 1643-1656, Aug. 2020.

\bibitem{4-JinkeRen}
J. Ren, G. Yu, Y. He, and G. Y. Li,
\newblock {``Collaborative cloud and edge computing for latency minimization,''}
\newblock {\em IEEE Trans. Veh. Technol.}, vol. 68, no. 5, pp. 5031-5044, May 2019.

\end{thebibliography}
\end{document}